\documentclass[chicago]{emulateapj}

\usepackage{amsmath}
\usepackage{amssymb}
\usepackage{graphicx}
\usepackage{wasysym}
\usepackage{multirow}
\usepackage{mdframed}

\slugcomment{Not to appear in Nonlearned J., 45.}

\shorttitle{Ammonia in High-mass Star Formation Regions}
\shortauthors{Lu et al.}

\begin{document}

\title{VLA observations of ammonia in high-mass star formation regions}
	
\author{Xing Lu\altaffilmark{1,2,5}} \author{Qizhou Zhang\altaffilmark{2}} \author{Hauyu Baobab Liu\altaffilmark{3}} \author{Junzhi Wang\altaffilmark{4}} \author{Qiusheng Gu\altaffilmark{1,5}}
\affil{$^1$School of Astronomy and Space Science, Nanjing University, Nanjing, Jiangsu 210093, China}
\affil{$^2$Harvard-Smithsonian Center for Astrophysics, 60 Garden Street, Cambridge, MA 02138}
\affil{$^3$Academia Sinica Institute of Astronomy and Astrophysics, P.O. Box 23-141, Taipei, 106 Taiwan}
\affil{$^4$Shanghai Astronomical Observatory, Chinese Academy of Sciences, Shanghai 200030, China}
\affil{$^5$Key Laboratory of Modern Astronomy and Astrophysics, Nanjing University, Nanjing 210093, China}

\begin{abstract}
We report systematic mapping observations of the NH$_{3}$ (1,1) and (2,2) inversion lines towards 62 high-mass star-forming regions using VLA in its D and DnC array configurations. The VLA images cover a spatial dynamic range from 40$''$ to 3$''$, allowing us to trace gas kinematics from $\sim$1 pc scales to $\lesssim$0.1 pc scales. Based on the NH$_3$ morphology and the infrared nebulosity on 1\,pc scales, we categorize three sub-classes in the sample: filaments, hot cores, and NH$_3$ dispersed sources. The ubiquitous gas filaments found on 1 pc scales have a typical width of $\sim$0.1\,pc and often contain regularly spaced fragments along the major axis. The spacing of the fragments and the column densities are consistent with the turbulent supported fragmentation of cylinders. Several sources show multiple filaments that converge toward a center, where the velocity field in the filaments is consistent with gas flows. We derive rotational temperature maps for the entire sample. For the three hot core sources, we find a projected radial temperature distribution that is best fitted by power-law indices from $-$0.18 to $-$0.35. We identify 174 velocity-coherent $\sim$0.1 pc scale dense cores from the entire sample. The mean physical properties for these cores are 1.1 km\,s$^{-1}$ in intrinsic linewidth, 18 K in NH$_{3}$ rotational temperature,  2.3$\times$10$^{15}$ cm$^{-2}$ in NH$_3$ gas column density, and 67 $M_{\odot}$ in molecular mass. The dense cores identified from the filamentary sources are closer to be virialized. Dense cores in the other two categories of sources appear to be dynamically unstable.
\end{abstract}


\section{INTRODUCTION}\label{introduction}

Interferometric observations found that high-mass stars form in $\lesssim$0.1 pc super-Jeans dense cores \citep{zhang2009, wang2011, csengeri2011}. How these high-mass star-forming cores are supported, and how they are related to the larger scale accretion flows, are being investigated extensively \citep{mckee2003, keto2010, baobab2012b, peretto2013}. When the high-mass protostars come into existence, the radiative and mechanical feedback further shape the cloud morphology and heat the gas, which can influence the subsequent formation of stellar clusters, and may eventually disperse the parental cloud \citep{churchwell2006,longmore2011,krumholz2014}. Systematic observations of the gas morphology, kinematics, and the temperature distribution of high-mass star-forming molecular clouds with a broad range of evolutionary stages, are therefore important to elaborate the life cycle of the dense molecular gas as well as the massive star clusters. 

The NH$_{3}$ inversion lines at $\nu\sim$24 GHz, which trace the critical density of $\sim$10$^4$ cm$^{-3}$ (Rohlfs \& Wilson 2004) and a broad range of gas excitation temperatures (10-50\,K for the (1,1), (2,2) and (3,3) lines;  Ho \& Townes 1983; Mangum et al. 1992), are one of the most useful molecular gas tracers for resolving the physical properties of star-forming gas. NH$_{3}$ has the advantage over the other molecular gas tracers (e.g. CO, CS; \citeauthor{bergin1997} \citeyear{bergin1997}) that it does not deplete efficiently onto the dust grain surface (Busquet et al. 2010).

NH$_3$ surveys toward high-mass star-forming regions have found signs of dependence of physical properties on stellar feedback. Studies of infrared dark clouds (IRDCs) \citep{pillai2006,ragan2011,chira2013}, where the earliest phases of high-mass star formation take place, revealed a mean kinetic temperature of $\lesssim$15 K, a mean linewidth of $\lesssim$2 km s$^{-1}$ in dense ($n$$\gtrsim$10$^5$ cm$^{-3}$) and massive (M$\gtrsim$10$^3$ M$_{\odot}$) IRDC clumps. Observations toward more evolved regions that show significant star formation via masers or infrared emission but have no signs of ultra-compact H{\sc ii} (UCH{\sc ii}) regions \citep{molinari1996,sridharan2002,wu2006,longmore2007,urquhart2011} found higher temperatures of $\sim$20 K, and linewidths of $\sim$2 km s$^{-1}$. Furthermore, studies of UCH{\sc ii} regions \citep{churchwell1990,harju1993,molinari1996,sridharan2002} revealed that most of them have temperatures above 25\,K, and linewidths of $\gtrsim$3 km s$^{-1}$. Therefore, as the cores evolve from starless to protostellar, the temperature as well as the non-thermal motions increase probably because of the feedback from protostars.

Although previous studies revealed global properties in high-mass star formation regions, most of the observations were carried out with single dish telescopes with an angular resolution of $>$$30''$, which corresponds to 0.5 pc at a distance of $\sim$3 kpc. Given that young massive stars typically form in local over-densities on $\lesssim$0.1\,pc scales \citep[e.g.][]{beuther2007,zhang2009,qiu2012}, the high angular resolution interferometric observations are necessary to characterize the physical properties of the gas accretion flow immediately around protostars (e.g. mean density, clumpiness, mass segregation), and to resolve the energy feedback \citep{baobab2012a,wang2012}. An example is a recent study by S\'{a}nchez-Monge et al. (2013) that reported 0.05 pc resolution National Radio Astronomy Observatory (NRAO) Very Large Array (VLA) NH$_{3}$ observations towards 15 intermediate-/high-mass star-forming regions. 

Here we present the VLA observations towards a sample of 62 high-mass star forming regions in NH$_3$ (J, K) = (1,1) and (2,2) inversion lines. Our observations covered 62 luminous ($L_{bol}>10^3\,L_{\odot}$) young massive clusters which are bright in far infrared ($F_{60\mu m}>50$-$100$ Jy) and show NH$_3$ detections in the previous Effelsberg 100m telescope observations  \citep{molinari1996,sridharan2002,wu2006}. The unprecedentedly large sample as compares with all previous interferometric surveys provides more sources in each of the morphological class or evolutionary stage, which improves the robustness in statistical comparisons. The paper is organized as follows. Details of the observations are given in Section \ref{observation}. We introduce the procedures to analyze the NH$_3$ data in Section \ref{reduction}. We classify NH$_3$ emission in classes of morphologies and analyze the relation between gas temperature and infrared emission, and discuss the distribution and correlation of physical properties in Section \ref{results}. Then in Section \ref{discussions} we explore several possible scenarios of the formation and fragmentation of filamentary sources as well as try to interpret the radial temperature distribution in the hot core sources. In Section \ref{conclusions}, we summarize the main results and present our conclusions.

\section{OBSERVATIONS AND DATA REDUCTION}\label{observation}

We observed 54 high-mass star-forming regions using the NRAO\footnote{The National Radio Astronomy Observatory is a facility of the National Science Foundation operated under cooperative agreement by Associated Universities, Inc.} VLA, in D/DnC array configurations during June 2004 (project code: AC0733), which constitute a good sample of high-mass protostars given the large luminosity and association with molecular outflows and dense NH$_3$ gas \citep{molinari1996,sridharan2002,beuther2002maser,wu2006}. The $uv$ distances of the observations range from 2\,k$\lambda$ to $\geqslant$81\,k$\lambda$, which yield $\lesssim$3\,arcsec angular resolution and $\sim$40\,arcsec maximal detectable angular scales. The correlator setup provided two independently tunable spectral windows with a 48.8 kHz spectral channel spacing and a 3.125 MHz bandwidth, corresponding to 0.618 km s$^{-1}$ channel width and 39 km\,s$^{-1}$ velocity coverage for the NH$_3$ (1,1) transition. These two spectra windows simultaneously covered the main and two inner satellite transitions of the NH$_3$ (1,1) line, as well as the main and one inner satellite transitions of the NH$_3$ (2,2) line.

The quasars 0137+331 (3C48) and 1331+305 (3C286) were used for flux calibration. Bandpass calibration was done with observations of 0319+415 (3C84), 1229+020 (3C273) and 2253+161 (3C454.3). Gain calibration was performed with observations of 0530+135, 1743$-$038, 1851+005, 1925+211 and 2322+509 in a cycle of 15 minutes. Phase calibrators were also used for the referencing pointings.

We also retrieved the archival VLA data towards another 8 star-forming regions observed in four tracks done in December 1997 (project code: AZ0103), July 2001 (project code: AS0708), January 2003 (project code: AS0747) and November 2005 (project code: AW0666). 

The data were calibrated and imaged (robust=0 weighted) using the NRAO Astronomical Image Processing System (AIPS). We did not perform the primary correction, in order to suppress the noise on the edge of the 2$'$ primary beam. This will lead to the underestimation of column densities and masses by a factor of $\sim$25\%, which is secondary to the uncertainties brought in by the distance and abundance ratio ambiguities. The synthesized beam of the NH$_3$ data is typically 3$''$$\times$3$''$, while we gridded the datacubes by a cell size of 0.8$''$. The rms noise level of these targets, except for IRAS 18264$-$1152 and IRAS 18517+0437 observed in projects AZ0103 and AS0708, is $\lesssim$3 mJy beam$^{-1}$ per 0.618 km s$^{-1}$ channel with a typical on-source time of 6 minutes. The observations are summarized in Table \ref{obssummary}. 

\section{NH$_3$ DATA ANALYSIS}\label{reduction}

The synthesized beams, the phase centers, and the velocity resolutions of the NH$_3$ (1,1) and (2,2) transitions are slightly different because of the different frequencies. Therefore we first regridded the NH$_3$ (2,2) data so that datacubes of the two NH$_3$ transitions have identical frames of coordinates and velocity.

\subsection{NH$_3$ Optical Depth and Gas Temperature}\label{reduction_1}

To improve the signal-to-noise ratio (SNR) in the NH$_3$ data, we performed a spatial and spectral smoothing. The spatial smoothing was done with a gaussian kernel of a full-width at half-maximum (FWHM) of 3$''$, comparable to the resolution, and the spectral smoothing was applied with a FWHM of 2\,km\,s$^{-1}$. The data were convolved with the gaussians, while the original cell size and channel width were preserved.

Among the 62 sources, 58 have sufficient SNR ($\geqslant$5 after smoothing) in the NH$_3$ (1,1) lines, 53 have SNR above 5 in both NH$_3$ (1,1) and (2,2) lines. For the sources with only NH$_3$ (1,1) detections, the rotational temperature cannot be derived. Table \ref{basicproperties} presents basic properties and NH$_3$ detections of the sample.

Following the procedure in \cite{Ho1983} and \cite{mangum1992}, we derived the optical depth of the (1,1) main hyperfine lines $\tau(1,1,m)$, assuming i). equal beam filling factor and excitation temperature for all hyperfine transitions ii). local thermodynamic equilibrium (LTE) and iii). that we could ignore the background temperature of 2.7 K,
\begin{equation}\label{tau11}
\frac{T_B(1,1,m)}{T_B(1,1,s)}=\frac{1-e^{-\tau(1,1,m)}}{1-e^{-a\tau(1,1,m)}},
\end{equation}
where $a=0.278$ is the relative intensity, $T_B(1,1,m)$ and $T_B(1,1,s)$ are the observed brightnesses of the main and the inner satellite lines. Note that when we computed $\tau(1,1,m)$ as well as $\tau(2,2,m)$ below, we set a lower limit of 0.001 and an upper limit of 10. They are not physical optical depths but rather for flagging purpose. For example, when we have a significant detection of the main hyperfine component of NH$_3$ (1,1), but not its satellite components, we assumed it to be optically thin and assigned a value of 0.001, which would be treated as optically thin in the calculation of temperature and column density.

We then derived $\tau(2,2,m)$, the optical depth of the (2,2) main hyperfine lines, under an additional assumption that all transitions under consideration have the same excitation temperature,
\begin{equation}\label{tau22}
\frac{T_B(1,1,m)}{T_B(2,2,m)}=\frac{1-e^{-\tau(1,1,m)}}{1-e^{-\tau(2,2,m)}}.
\end{equation}

Finally we obtained the rotational temperature $T_R$ from:
\begin{equation}\label{trotation}
T_R=-\frac{41.5\ {\rm K}}{\ln[0.282\frac{\tau(2,2,m)}{\tau(1,1,m)}]}.
\end{equation}
It is worth noting that due to the variation of the denominator in Equation \ref{trotation}, the rotational temperature is not sensitive to the optical depth ratio when $T_R\gtrsim30$ K. As a result, rotational temperatures cannot be reliably determined using the NH$_3$ (1,1) and (2,2) transitions when they are above $\sim$30 K. This limitation arises from the fact that the two NH$_3$ transitions have a relatively low energy levels of 23 and 65 K, respectively, thus are not sensitive to the warmer gas. Data from higher (J,K) transitions alleviate this limitation \citep[]{zhang1999}. On the other hand, to minimize the contribution from the cosmic microwave background (CMB) emission, we set a lower limit of 8 K (3 times the CMB temperature 2.7 K) for the rotational temperature.

Alternatively, when the optical depth derived from Equation \ref{tau11} is smaller than 0.5, including when it is flagged as 0.001, we considered it to be optically thin, and derived the temperature directly from the ratio of brightnesses, e.g.
\begin{equation}\label{trotation_thin}
T_R=-\frac{41.5\ {\rm K}}{\ln[0.282\frac{T_B(2,2,m)}{T_B(1,1,m)}]}.
\end{equation}

To calculate the kinetic temperature $T_K$, we applied the empirical relationship in \cite{walmsley1983}:
\begin{equation}
T_R=\frac{T_K}{1+\frac{T_K}{41.5}\ln[1+0.6\exp(-\frac{15.7}{T_K})]}.
\end{equation}
When $T_R\lesssim30$K, the kinetic temperature increases approximately linearly with $T_R$, therefore we can use $T_R$ as an equivalence of $T_k$ to trace the thermal states, as what we did in Figures \ref{tempfigs} \& \ref{tempfigs_more}. When $T_R\gtrsim30$ K, or the corresponding $T_K\gtrsim40$ K, the linear dependency of $T_R$ on $T_K$ breaks down quickly so that $T_R$ is no longer a reliable indicator of gas temperature but a lower limit. 

\subsection{Intrinsic Linewidth}\label{reduction_2}

To derive the intrinsic linewidth, we first subtracted the channel width of 0.618 km s$^{-1}$ quadratically from the FWHM linewidth, the latter is measured from the second-order moment of the NH$_3$ (1,1) main hyperfine lines before the spatial and spectral smoothing. We rejected the pixels where the original linewidth is smaller than the channel width, and those where the linewidth after the subtraction is smaller than the thermal linewidth derived from the rotational temperature. 

The NH$_3$ main and satellite inversion lines consist finer hyperfine splittings which is smaller than the channel spacing in our VLA observations. In addition, the optical depths of the NH$_3$ (1,1) line transitions are usually $>$1. Therefore to retrieve the intrinsic linewidth, we followed \cite{barranco1998} to generated a lookup table with series of NH$_3$ optical depths $\tau(1,1,m)$ and intrinsic linewidths $\Delta v$. The relation between the input parameters and the observed spectrum is: 
\begin{equation}\label{intrinsiclw}
\begin{split}
T_{B}(1,1,m)^{(v)}&=T_{ex}\left\{ 1-\exp\left[-\tau^{(v)}\right]\right\}, \\
\tau^{(v)}=\tau(1,1,m)\displaystyle\sum_{j=1}^{8}\alpha_j&\exp\left\{ -4\ln2\left[\frac{v-(v_{LSR}+v_j)}{\Delta v}\right]^2\right\},
\end{split}
\end{equation}
where $T_{ex}$ is the effective excitation temperature, $\alpha_j$ is the relative intensity, $v_j$ is the velocity of the $j$th hyperfine component with respect to the central velocity $v_{LSR}$, and $\Delta v$ is the intrinsic FWHM linewidth of the hyperfine components.  The above relation links the blended linewidth measured from the computed spectrum with an intrinsic linewidth and optical depth. Therefore, one can trace back to the intrinsic linewidth given the optical depth and the blended linewidth.

\subsection{Column Density and Mass}\label{reduction_3}

To evaluate the NH$_3$ column density, first we assumed the excitation temperature of the (1,1) transition $T_{ex}(1,1)=T_R$, which is valid under the LTE assumption. Then the column density of the (1,1) transition follows \citep{mangum1992}:
\begin{equation}\label{column}
N(1,1)=6.60\times10^{14}\frac{T_R}{\nu(1,1)}\tau(1,1,m)\Delta v\ {\rm cm^{-2}},
\end{equation}
where $\nu(1,1)$ is the transition frequency in GHz, and $\Delta v$ is the intrinsic FWHM linewidth in km s$^{-1}$ (see Section \ref{reduction_2}); or when it is optically thin ($\tau(1,1,m)<0.5$),
\begin{equation}\label{column_thin}
N(1,1)=6.60\times10^{14}\frac{T_B(1,1,m)}{\nu(1,1)}\Delta v\ {\rm cm^{-2}}.
\end{equation}
Note that for Equation \ref{column_thin} we assumed the beam filling factor to be unity, which might underestimate the column density by ~20\% given that the filling factor of similar NH$_3$ cores is usually 0.6-0.8 \citep[e.g.][]{mangum1992}.

For the sources with only NH$_3$ (1,1) detections, we assumed optically thin (1,1) emission, therefore the column density was also derived by Equation \ref{column_thin}.

The total NH$_3$ column density $N(\rm NH_3)$ can be derived as in \cite{rohlfs2004}:
\begin{equation}\label{totalcolumn}
\begin{aligned}
N(\rm NH_3)=\frac{1}{3} N(1,1)Q_{rot}\exp(\frac{23.1}{T_R})\\
\approx0.0138\times N(1,1)\exp(\frac{23.1}{T_R})T_R^{3/2}.
\end{aligned}
\end{equation}
Here, $Q_{rot}$ is the partition function for which we used the approximation $Q_{rot}\approx168.7\sqrt{\frac{T_R^3}{B^2C}}$, the rotational constants $B$=$298.117$ and $C$=$186.726$ in GHz. This approximation leads to an error smaller than 10\% in the range of $T_R=$10-50 K.

Assuming a uniform NH$_3$ fractional abundance [NH$_3$]/[H$_2$]=$3\times10^{-8}$ \citep{harju1993}, we obtained the molecular mass $M(\rm H_2)$ by summing up $N(\rm NH_3)$ within the area. 

\subsection{Uncertainties of the Physical Properties}\label{error}

First, we considered the uncertainty of the rotational temperature. The calibration errors do not affect the optical depth hence the temperature because they are cancelled in the ratios of brightness temperatures. Missing fluxes in the VLA observations can affect the rotational temperature. \cite{wang2007} compared the result of their VLA observations of spatially extended clouds with single-dish data and found that the missing flux is equivalent to an intensity of 3 mJy beam$^{-1}$, comparable to the rms level of the NH$_3$ data. Since we used a threshold of 5$\times$rms when deriving the temperature, the effect of missing flux is negligible.

The algorithm we used in Section \ref{reduction_1} also introduces uncertainty into the temperature. According to \cite{li2003}, the rms of the derived temperature is approximately $\frac{10.3}{\rm SNR}$ K where SNR is the signal-to-noise ratio of the NH$_3$ (2,2) lines. Since we have set a criterion of SNR$\geqslant$5, the SNR ranges from 5 to $\gtrsim$20. Therefore the uncertainty in the temperature due to the algorithm is about 2 K in the envelopes where the SNR$\sim$5 and $\lesssim$0.5 K at the peaks where the SNR is much higher.

In general, we expected an uncertainty of $\pm$2 K at most in the rotational temperature, while for regions we were interested the SNR is always large enough to ensure $\lesssim$1 K uncertainty.

For the intrinsic linewidth $\Delta v$, the uncertainty depends on both the optical depth and the blended linewidth, as indicated in Equation \ref{intrinsiclw}. The optical depth, decided by the ratios of brightnesses, is to the same precision as the temperature, therefore the majority of the uncertainty lies in the blended linewidth. Since we measured the latter from the second-order moment maps, the uncertainty of linewidth has a dependency of $$\frac{\delta\Delta v}{\Delta v}=\frac{1}{2}\frac{\delta F}{F},$$ where $F$ is the integrated intensity. As we estimated the uncertainty of $F$ to be 40-50\%, the uncertainty of the intrinsic linewidth therefore is 20\%.

For properties such as column density and molecular mass, the error in calibration will be inherited and leads to an uncertainty of $\sim$20 \%. The systematic uncertainty of the column density can be even larger considering the bias of the missing flux, the beam filling factor, and the abundance ratio [NH$_3$]/[H$_2$]. For the mass, the error of the distances brought in even more uncertainty. For example, after comparing the trigonometric parallax distance and the kinematic distance of IRAS 18159$-$1648 (1.12 kpc vs. 2.46 kpc; \cite{sato2010}), we found that the kinematic distance is in error by a factor of 2, suggesting that the mass based on such kinematic distances could be largely biased. Because of the reasons enumerated above, we expect large uncertainties (at least $\pm$50 \%) in the column density, and even larger uncertainties and biases in the mass.

In order to examine the robustness of the mass, first, we compared our result with those derived from single-dish dust continuum observations by \cite{beuther2002cs}. In Table \ref{physical}, 35 sources overlap with the dust continuum sample. We scaled their masses with the distances listed in Table \ref{basicproperties} and attached them in Table \ref{physical}, noted by $M_d$. The largest $M({\rm H_2})/M_d$ ratio is $\sim$2, found in IRAS 18337$-$0743. The smallest ratio is down to $<$1/10. Considering the uncertainty factor of $\gtrsim$5 in $M_d$ due to dust emissivity, temperature and gas-to-dust ratio, as well as the uncertainty in $M({\rm H_2})$, the two masses are fairly consistent on 1 pc scales. Second, we obtained the SMA dust emission data at 0.1 pc linear resolution of one hot core and several filamentary sources  (see Section \ref{morphology}), and compared core mass derived from dust with our results. For example, assuming a dust temperature of 25 K, a gas-to-dust ratio of 100, and an emissivity index of 1.5, the mass of the hot core IRAS 18182$-$1433 based on dust emission is 448 M$_\odot$, while that based on NH$_3$ data is 264 M$_\odot$. On the other hand, the mass of the central core of the filamentary source IRAS 18308$-$0841 based on dust emission is 136 M$_\odot$, while that based on NH$_3$ data is 373 M$_\odot$. It is likely that the abundance ratio of $3\times10^{-8}$ is too large for hot cores, and too small for filamentary sources. However we have only 7 sources that have been observed by the SMA, out of which 5 are filaments, 1 is a hot core, and 1 is irregular. Before we have more high angular resolution dust emission observations for comparison, we will keep applying a uniform abundance ratio for all the sources. As a result, the core masses and virial parameters in Figures \ref{histogram} \& \ref{histo_groups} and Table \ref{fullcorelist} must be treated with caution.

\section{RESULTS}\label{results}

The observational properties of the sample are listed in Table \ref{basicproperties}. The coordinates in equatorial and galactic systems mark phase centers of NH$_3$ (1,1), and the velocity corresponds to the V$_{LSR}$ where the peak emission is found. Except those otherwise noted, all the distances were calculated using the Galactic rotation model of \cite{Reid2009}, with the coordinates and velocity in Table \ref{basicproperties}, assuming that the orbital speed of the Sun $v$=$254$ km s$^{-1}$ and the distance from the Sun to the Galactic center $d$=$8.4$ kpc. Then we used the flux of all four bands of IRAS to calculate a luminosity, following \cite{sanders1996}. Note that the angular resolution of IRAS is low (0.5-2 arcmin, depending on the wavelengths) so the luminosity we derived is that of the entire region.

The kinematic distance ambiguity cannot be completely resolved at present. However we attempted for some of them with the empirical relation of \cite{fish2003}, to confirm their near kinematic distance. The near kinematic distance is also preferred if the source is recognized as an IRDC candidate. We also resorted to associations with stars, H{\sc ii} regions or masers to obtain a more reliable distance, based on the references listed in Table \ref{basicproperties}. For remaining sources with distance ambiguities, the near kinematic distance was applied.

The derived physical properties were given in Table \ref{physical}. The linewidth, optical depth, rotational and kinetic temperatures, column density and mass were derived following the methods described in Sections \ref{reduction_1} to \ref{reduction_3}, and the ranges of values within the considered regions were listed. 

\begin{figure}[!t]
\begin{tabular}{p{9.0cm}p{8.1cm}}
\includegraphics[width=9.0cm]{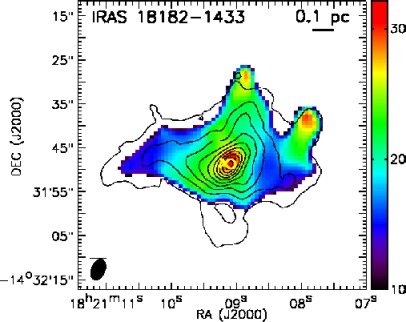} & \includegraphics[width=8.1cm]{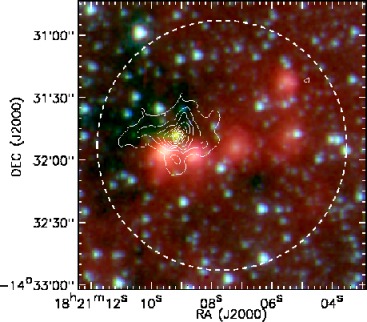} \\
\includegraphics[width=9.0cm]{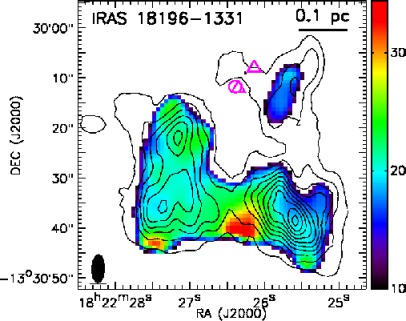} & \includegraphics[width=8.1cm]{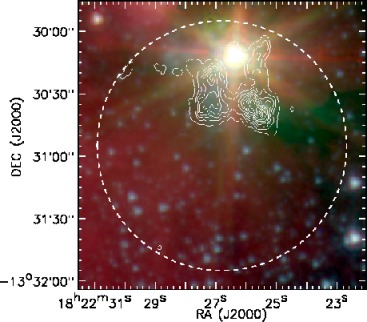} \\
\includegraphics[width=9.0cm]{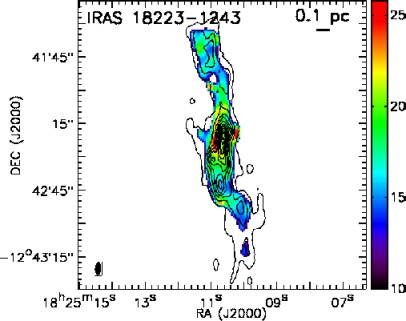} & \includegraphics[width=8.1cm]{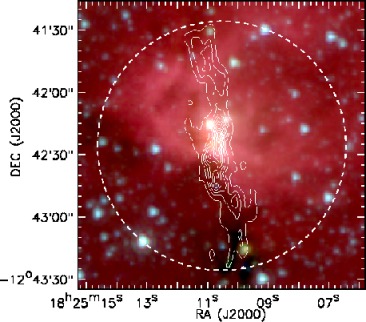} \\
\end{tabular}
\caption{Rotational temperature and IRAC three-color images of three representative sources: from top to bottom, a hot core source, a NH$_3$ dispersed source, and a filamentary source. More images are in the appendix. {\it Left}: Rotational temperatures are presented in color scales. The black contours are the integrated intensities of NH$_3$ (1,1) main hyperfine line, from 5\% to 95\% of the peak intensity in steps of 10\%. Maser detections with interferometers are marked with either circles (water maser), crosses (Class II methanol maser), or triangles (hydroxyl maser). The synthesized beam of the integrated intensities is shown in the lower-left corner in each panel, while the beam of the rotational temperatures should be slightly larger given that the temperatures are derived from the smoothed data. {\it Right}: The Spitzer IRAC composite image, with 8 $\mu$m (red), 4.5 $\mu$m (green) and 3.6 $\mu$m (blue) emission. The NH$_3$ (1,1) integrated intensity contours are identical to those in the left panels. The dashed circle shows the 2$'$ FWHM primary beam of the VLA.}
\label{tempfigs}
\end{figure}

We presented the rotational temperature distribution in Figures \ref{tempfigs} \& \ref{tempfigs_more}. Next to the temperature images are the three-color infrared images made with 3.6/5.8/8.0 $\mu$m bands of {\it Spitzer} IRAC \citep{werner2004,fazio2004}. For IRAS 05490+2658 and IRAS 06061+2151, only 3.6 and 5.8 $\mu$m bands data are available. There are also four sources (IRAS 20278+3521, IRAS 23033+5951, IRAS 23139+5939, IRAS 23545+6508) that are not covered by any IRAC observation, so we presented the MSX images \citep{price2001} comprised of three bands centered at 8.28, 12.13 and 14.65 $\mu$m. All the images are overlaid with contours of the integrated intensity of the NH$_3$ (1,1) main hyperfine line, starting from 5 \% of the peak integrated intensity and in step of 10 \%. We also found available maser detections from the literature by either single-dish or interferometric observations and listed them in Table \ref{basicproperties}. However only high angular resolution interferometric results were marked with corresponding symbols in Figures \ref{tempfigs} \& \ref{tempfigs_more}.

\subsection{Morphology of NH$_3$ Emission}\label{morphology}

We resolved high-mass star-forming regions into different morphological classes: the filaments, the relatively symmetric gas concentrations, and the more irregularly shaped extended envelopes, which are introduced in the following sections. Typical morphological classes such as filaments, hot cores and NH$_3$ dispersed sources are marked in Table \ref{physical}.

\subsubsection{Filaments}\label{result_filaments}

In our sample, we referred to NH$_3$ emission with a large aspect ratio (length/width$\gtrsim$3) as a ``filament''.
We found that filaments are universal. For example, IRAS 18223$-$1243 is filamentary in the north-south direction, while sources such as IRAS 18308$-$0841 show an irregular geometry on $>$1\,pc scales, but are embedded with filamentary substructures. 

One common feature of these sources is the NH$_3$ peaks along the filaments, as expected from the cylindrical collapse models \citep{chandra1953, nagasawa1987}. The fragments are usually regularly spaced and coherent in velocity. For example, IRAS 19074+0752 has four peaks within the long filament in the southwest, as shown in Figure \ref{tempfigs_more}. The spatial interval between the peaks is $\sim$6 arcsec, and the velocity dispersion is less than 1.2 km s$^{-1}$. It is interesting that these four emission peaks do not coincide with temperature peaks; some of them do not even have enough NH$_3$ (2,2) SNR to derive a valid temperature. The high temperature regions are preferentially located in between these peaks. This trend is still valid even considering that the SNR of the NH$_3$ emission between the peaks is lower than that in the peaks thus the uncertainty of the temperature is larger ($\pm$5 K for most of them).

Velocity gradients are also seen in the filaments, most of time along the major axis. While some cases, such as IRAS 18223$-$1243, show an ordered, continuous velocity gradient along the filaments, it does not necessarily represent systematic motions. For example, the filamentary part extending to the south in IRAS 18507+0121 has two distinct velocity components, which are more likely two independent structures lying close in the light of sight.

In IRAS 18308$-$0841, IRAS 18310$-$0825, IRAS 18337$-$0743, IRAS 18460$-$0307, and IRAS 18472$-$0022, the filaments appear to spatially converge to a common center. This geometry resembles the``hub-filament system'' (HFS) introduced in the previous survey of nearby young low-mass star clusters \citep{myers2009,myers2011} as well as a few high-mass cases \citep{galvan2010,galvan2013,baobab2011,baobab2012a,baobab2012b,peretto2013}. The five HFS sources have unique morphologies: the filaments are often associated with the infrared dark regions, indicating a high column density, and the hubs have bright infrared counterparts. IRAS 18460$-$0307, for example, contains four major filaments. The ones extending to the west and the south are associated with infrared extinctions, as shown in Figure \ref{tempfigs_more}, consistent with the high column density of $\gtrsim$10$^{23}$ cm$^{-2}$ derived from NH$_3$ emission. There is a bright infrared source at the hub, where the NH$_3$ emission is not present. 

\begin{figure}[!t]
\centering
\includegraphics[width=12.0cm]{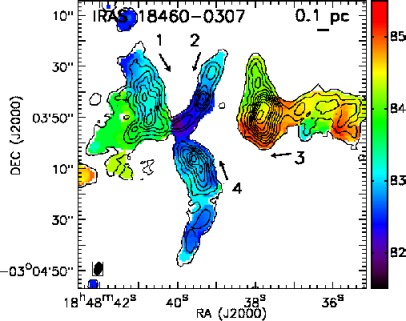}
\caption{The velocity field of IRAS 18460$-$0307. The contours represent the integrated NH$_3$ (1,1) emission plotted in steps of 10\% starting from 5\% of the peak flux. Potential converging filaments are marked by arrows.}
\label{18460_mom1}
\end{figure}

The kinematics of the filaments in HFS appears to be well-defined. Again we took IRAS 18460$-$0307 as an example: as presented in Figure \ref{18460_mom1}, the four filaments, marked by arrows, are either red or blue shifted with respect to the hub, revealing plausible converging flows of gas toward the center. 

\subsubsection{Concentrations}

By concentration, we referred to NH$_3$ structures that have a high SNR ($\geqslant$10) at the center and a roughly circular envelope around. In addition, we required these sources not to be embedded in resolved filaments.

Representatives of this morphology include IRAS 18089$-$1732, IRAS 18182$-$1433, IRAS 18360$-$0537 and IRAS 18414$-$0339. They all have one dominant emission peak where the temperature rises to $\gtrsim$30 K. By contrast, the temperature in the envelope is mostly $\lesssim$20 K. However in IRAS 18360$-$0537 and IRAS 18414$-$0339 we also found temperature elevation in the envelope up to 30 K. It can be heated by nearby stars or outflows from embedded protostars \citep[e.g.][]{zhang1999, zhang2002}, but additional corroborating evidence is needed to differentiate the two possibilities.

\begin{figure}
\centering
\includegraphics[width=12.0cm]{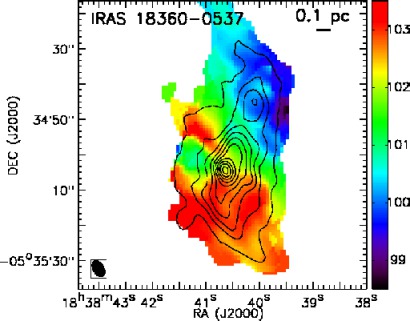}
\caption{The velocity field of IRAS 18360$-$0537. The contours represent the integrated NH$_3$ (1,1) emission plotted in steps of 10\% starting from 5\% of the peak flux. }
\label{18360_mom1}
\end{figure}

The kinematics in the envelope shows signatures of rotation around the peak. In IRAS 18360$-$0537 for example, the envelope is extended to the north and south. The northern part is blue-shifted by 2 km s$^{-1}$ with respect to the peak, while the southern part is almost symmetrically red-shifted, shown in Figure \ref{18360_mom1} \citep[cf.][Figure 18]{lijuan2012}.  This gradient is consistent with that reported by \citet{qiu2012} based on the SMA observations. 

In all the concentrated sources except IRAS 18414$-$0339, we found a very broad linewidth at the NH$_3$ emission peak. The linewidth can be up to $\gtrsim$5 km s$^{-1}$ within $\sim$3 arcsec, and can be attributed to rotation around a massive protostar \citep{zhang1998,zhang2002,zhang2007}. \citet{keto2010} compared observations towards the high-mass star forming region IRAS 20126+4104 against model spectra from an envelope in Ulrich flow \citep{ulrich1976} with an accretion disk, and concluded that infall and rotation motions are able to reproduce the large linewidth. Indeed, the inteferometric sub-mm observations revealed signature of rotation and infall, as well as abundant organic molecular lines and outflows in the three sources \citep{beuther2004a,beuther2004b,beuther2006,qiu2012}, indicating that they are star-forming hot cores with embedded rotating disks or envelopes. 

Hereafter we only focus on the three ``hot core'' sources when referring to the concentrations.
 
\subsubsection{Other Concerned Sources}

We found that in some sources the NH$_3$ emission surrounds an infrared source, but it is not detected at the IR source position. IRAS 06382+0939, for example, has a shell shaped NH$_3$ emission around an infrared source, as shown in Figure \ref{tempfigs_more}. The luminosity of the IRAS source is 440 $L_{\odot}$. Another case is IRAS 18196$-$1331, where several pieces of NH$_3$ emission enclose a bright infrared source with a luminosity of 6.3$\times$10$^4\ L_{\odot}$. The projected distance of the NH$_3$ peaks from the central infrared sources is 0.1-0.2 pc, comparable to the typical size of a star-forming dense core. \citet{zinchenko1997} found similar NH$_3$ underabundance in clouds associated with very luminous ($\sim$10$^5\ L_{\odot}$) infrared sources. They suggested that the strong radiation field from luminous protostars increases the abundance of C$^+$ which destroys NH$_3$. Neither of our sources is as luminous as this scenario requires, however the chemical equilibrium with carbon based molecules can have an influence on the NH$_3$ underabundance. We compared the physical properties in these sources with those of spatially concentrated and filamentary sources, and presented the result in Section \ref{3classes}. It turns out that their properties are quite similar to those of the cores found in filamentary sources which are quiescent and cold, indicating a low level of star formation.

\subsection{Association of Temperature Distribution and Infrared Sources}\label{infrared}

Infrared emission in 8 $\mu$m traces the hot dust around the massive stars and the spatially extended polycyclic aromatic hydrocarbons (PAHs) line features excited in UV photon illuminated regions \cite[e.g.][]{peeters2011}. The systematic comparison between the temperature distribution and infrared emission helps to understand the heating of protostars upon their surrounding dense gas. 

\subsubsection{Embedded Heating Sources}
We found many cases in which the temperature peaks coincide with infrared sources. The infrared counterpart can be a group of sources, as in the case of IRAS 18159$-$1550, or a single source as in IRAS 06103+1523. The rotation temperature usually increases to $\gtrsim$30 K as compared to $\sim$20 K in the periphery regions. However in some sources high-temperature regions as well as their infrared counterparts are not associated with NH$_3$ emission peaks. IRAS 18159$-$1550 is such a case: it exhibits a hot ridge (T$_R$$\gtrsim$25 K) with three embedded infrared sources, among the cold NH$_3$ peaks (T$_R$$\lesssim$20 K), shown in Figure \ref{tempfigs_more}. No infrared sources were found in the NH$_3$ emission peaks. 

Nonetheless, most of the sources show a good agreement of temperature peaks, infrared sources and NH$_3$ emission peaks. In particular, we found some sources with an ordered temperature profile: a  warm core at the center and a much cooler envelope, with a temperature gradient. Usually they are typical ``hot molecular cores'' internally heated by one or a group of high-mass protostars. If we further assumed that the embedded protostars are closely packed so that the anisotropy in radiative heating can be neglected, then we can derive the projected radial temperature distribution $T(r_{prj})\propto r_{prj}^{-\alpha}$, where $r_{prj}$ is the projected distance from the center \citep[e.g.][]{scoville1976,garay1999,keto2010}.

We took three concentrated sources with such temperature gradient for analysis: IRAS 18089$-$1732, IRAS 18182$-$1433 and IRAS 18360$-$0537. Their temperature peaks are associated with NH$_3$ emission, indicating heating sources embedded in dense gas. For comparison, we also include IRAS 18488+0000 in the analysis, which also exhibits a radial temperature gradient but the temperature peak is somewhat offset from that of NH$_3$ emission. Note that in IRAS 18182$-$1433 there are two velocity components at $\sim$58 and 61 km s$^{-1}$ respectively. To extract a meaningful temperature distribution, we took the one at 58 km s$^{-1}$ because its NH$_3$ emission is more concentrated around the dust continuum peak found by \cite{beuther2006}.

\begin{figure}
\begin{tabular}{p{8.5cm}p{8.5cm}}
\includegraphics[width=0.45\textwidth]{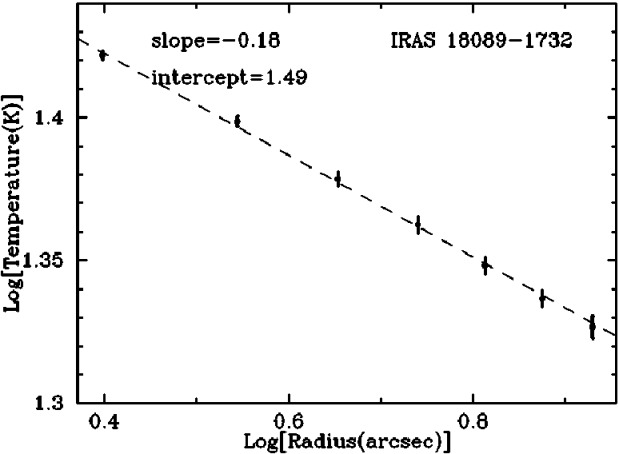} &\ \ \includegraphics[width=0.45\textwidth]{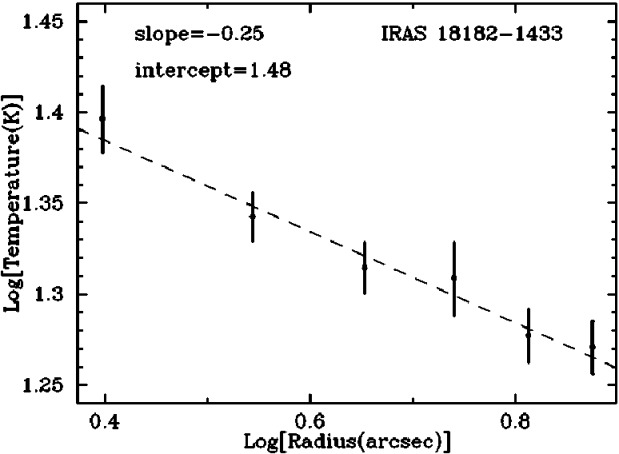} \\
\includegraphics[width=0.45\textwidth]{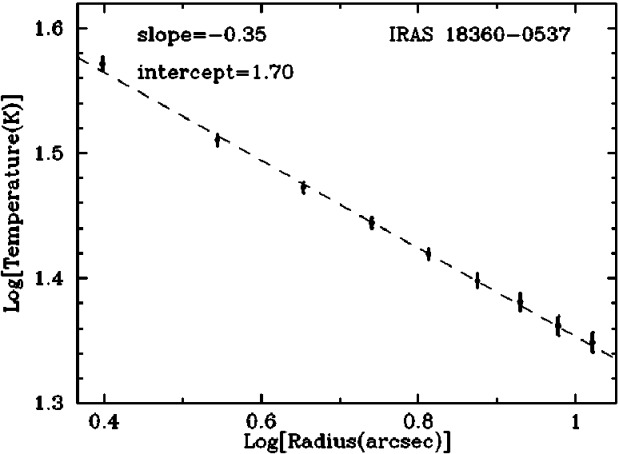} &\ \ \includegraphics[width=0.45\textwidth]{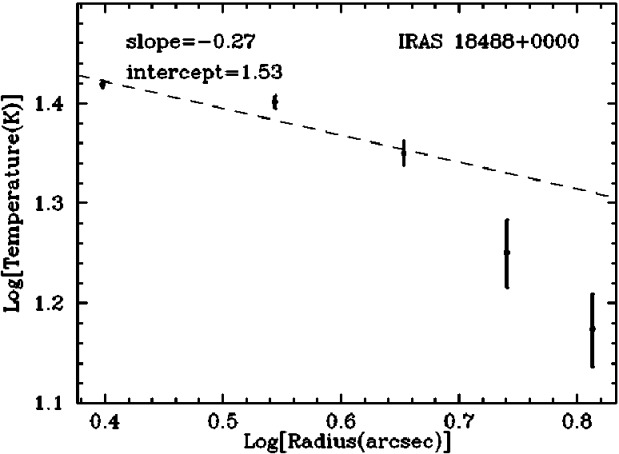} \\
\end{tabular}
\caption{Least squares fitting of temperature-radius. Error bars show the standard deviation of the mean in each radius bin.}
\label{tempgra}
\end{figure}

\begin{figure}
\centering
\begin{tabular}{p{8.5cm}p{8.5cm}}
\includegraphics[width=0.47\textwidth]{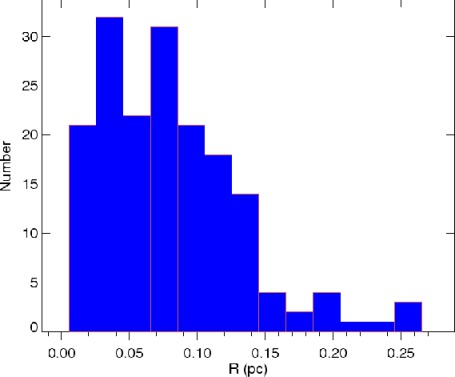} &\includegraphics[width=0.47\textwidth]{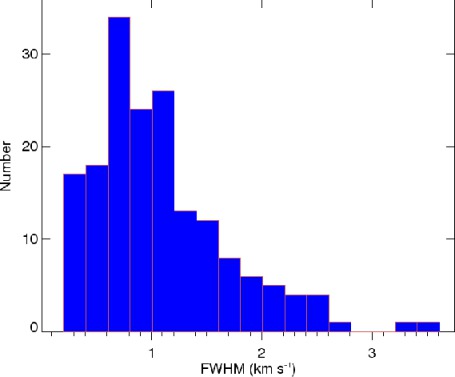} \\
\includegraphics[width=0.47\textwidth]{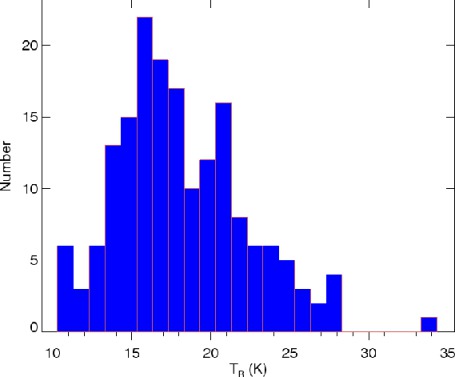} &\includegraphics[width=0.47\textwidth]{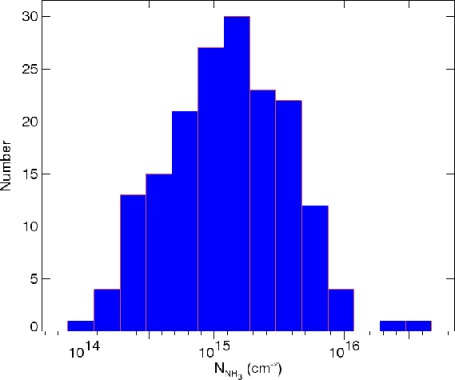} \\
\includegraphics[width=0.47\textwidth]{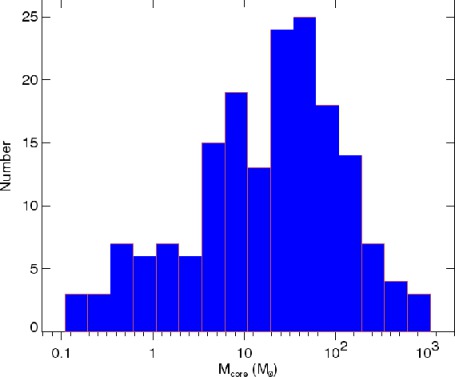} &\includegraphics[width=0.47\textwidth]{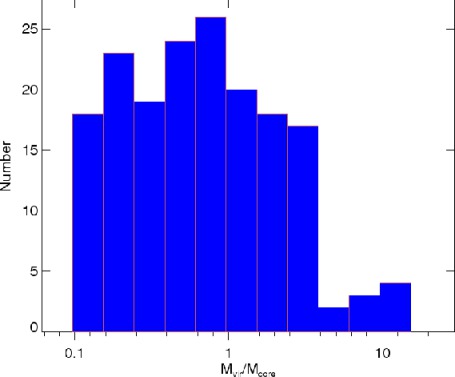} \\
\end{tabular}
\caption{Distributions of the overall core properties.}
\label{histogram}
\end{figure}

\begin{figure}
\centering
\begin{tabular}{p{8.5cm}p{8.5cm}}
\includegraphics[width=0.47\textwidth]{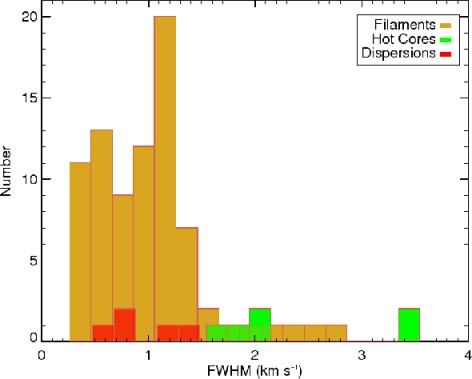} &\includegraphics[width=0.47\textwidth]{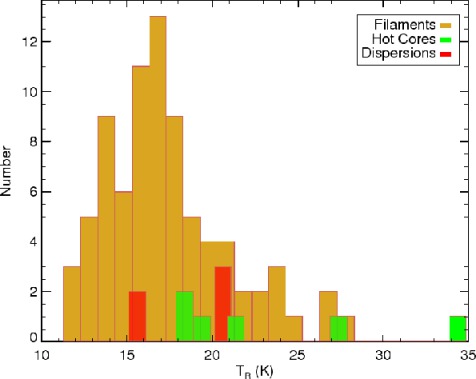} \\
\includegraphics[width=0.47\textwidth]{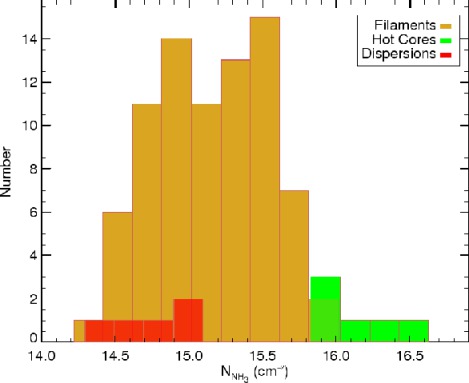} &\includegraphics[width=0.47\textwidth]{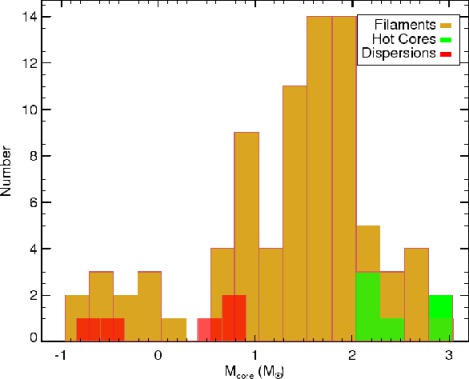} \\
\includegraphics[width=0.48\textwidth]{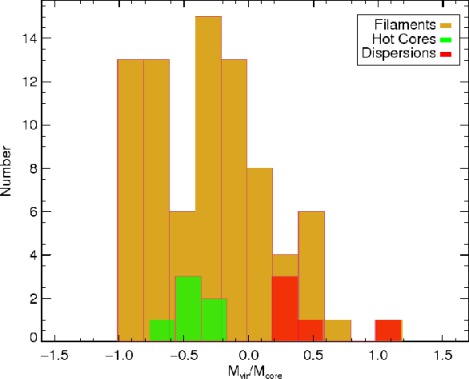} & \\
\end{tabular}
\caption{Distributions of the core properties, for the three classes.}
\label{histo_groups}
\end{figure}

We calculated the mean temperature concentrically averaged with respect to the temperature peak. The mean values versus the projected radii $r_{prj}$ and the fitting results are shown in Figure \ref{tempgra}. 

Following the discussion in \cite{longmore2011}, we fitted the power-law index only at the $\gtrsim$0.05 pc projected radius. This helps alleviate the bias caused by the insufficiently high energy level transitions among the observed NH$_3$ lines, and the smearing effects by the synthesized beam (see Section \ref{reduction_1}). The outer boundary of the fitted regions was chosen where the SNR of the NH$_3$ (2,2) line drops to 10.

For IRAS 18182$-$1433, we derived a power-law slope of $-$0.25$\pm$0.04 and an intercept of 1.48$\pm$0.03 in the logarithmic diagram using linear least-squares method, leading to the relation $T(r_{prj})=(30.2\pm2.0)\ \left(\displaystyle\frac{r_{prj}}{\rm arcsec}\right)^{-0.25\pm0.04}$ K. The uncertainty we used in the linear fitting is the standard deviation of the mean value, without considering the error propagated from temperature derivation as we discussed in Section \ref{error}. However inside the radius where the fitting was performed, the SNR in the NH$_3$ spectra is always larger than 10, mostly around 20, thus the uncertainty in rotational temperature itself is smaller than 1 K. Taking it into consideration will enlarge the uncertainty in the power-law index to $\pm0.05$, but will not change the slope. Similarly, we found $T(r_{prj})=(30.9\pm0.2)\ \left(\displaystyle\frac{r_{prj}}{\rm arcsec}\right)^{-0.18\pm0.02}$ K for IRAS 18089$-$1732, and $T(r_{prj})=(50.1\pm0.9)\ \left(\displaystyle\frac{r_{prj}}{\rm arcsec}\right)^{-0.35\pm0.02}$ K for IRAS 18360$-$0537. The slopes and intercepts were labeled in Figure \ref{tempgra}.

Note that the power-law relation we derived is based on the projected radius $r_{prj}$, instead of the real radius $r$. To reconstruct the relation between the temperature $T(r)$ and the radius $r$, it is necessary to decompose contributions to the observed temperature $T(r_{prj})$ from different optical depths. Radio transfer modeling will help to resolve this ambiguity \citep[e.g.][]{longmore2011}.

The additional sample IRAS 18488+0000 yields a projected power-law slope of $-0.27$, which translates to $T(r_{prj})\propto r_{prj}^{-0.27}$. However, the fit is poor with a goodness of fit of $\sim$0.003. The asymmetry inside the source may lead to the divergence from a well-defined radial temperature distribution: the protostars manifested by the masers may be surrounded by anisotropic medium thus the gas is heated in an anisotropic way. In addition, the external factor may also play a role in shaping its temperature structure: the infrared bubble to the west of the NH$_3$ emission peak, one of those catalogued in \cite{churchwell2006}, is associated with an H{\sc ii} region \citep{anderson2009}. The temperature gradient could be greatly altered by it, either by shocks or by UV radiation from the stars in it.

\subsubsection{External Heating Sources}

In many sources we found higher temperature toward the edge of the NH$_3$ emission than at the center. The observed trends of temperature distribution are significant even after considering the uncertainties introduced in Section \ref{error}. One example is IRAS 18272$-$1217. The NH$_3$ emission extends across 50 arcsec with an position angle of $\sim$10$^\circ$. At the northern end, the rotation temperature increases to $\gtrsim$20 K along the rim, while at the center it is around 15 K. According to Section \ref{error}, at the rim the uncertainty in temperature is $\sim$2 K. The uncertainty at the center is much smaller, normally less than 0.5 K thanks to the large SNR. Therefore the difference in temperature is valid. Furthermore, we find an infrared source to the northeast of this rim, indicating the presence of external sources whose radiation may explain the temperature elevation. 

\subsection{Dense Core Properties}\label{densecore}

\subsubsection{Dense Core Identifications and Overall Core Properties}\label{coreoverall}

We used the {\it clumpfind} algorithm \citep{williams1994} to search for dense cores in the NH$_3$ integrated intensity maps. In our searching process the lowest contour level was 20 mJy beam$^{-1}$ km s$^{-1}$, which is equivalent to $\sim$5$\sigma$ noise level of the integrated intensity maps; the contour increment was 10 mJy beam$^{-1}$ km s$^{-1}$. We rejected the structures on the boundary of the maps or discontinuity in space by visually inspection. We also checked the NH$_3$ data in position-position-velocity space to reject the false integrated intensity peaks due to the superimposed multiple velocity components along the line of sight. Because we would compare temperatures of cores, we also rejected those with so weak NH$_3$ (2,2) emission that we were not able to derive a temperature. In the end we found 174 dense gas cores. All the cores, including those rejected because of possible multiple velocity components and absence of temperatures, are listed in Table \ref{fullcorelist} in the appendix.

To estimate the intrinsic core radius $R$, we first derived the effective half-maximum radius $R_{eff}$, i.e. the radius of the circle that covers the same area within the half-maximum contour in the integrated intensity maps. Then we subtracted the synthesized beam (FWHM major and minor axes $bmaj$ and $bmin$) from it using: $R=\sqrt{R_{eff}^{2}-\frac{1}{4\ln2}bmaj\times bmin}$.

The intrinsic linewidth $\Delta v$, the rotational temperature $T_R$ and the column density $N({\rm NH_{3}})$ were averaged within the effective radius. The molecular mass $M_{core}$ was considered within the same area. With the radius R and the intrinsic velocity dispersion $\sigma_v=\Delta v/(2\sqrt{2\ln2})$ we were able to obtain the virial mass \citep{maclaren1988}, assuming a uniform radial density distribution: $$M_{vir}=\frac{5R\sigma_v^2}{G}.$$

The mean and median values of the core properties are listed in Table \ref{stat}. The radius is $\sim$0.08 pc, consistent with the typical size of the dense cores \citep{jijina1999}. The mean FWHM linewidth is 1.08 km s$^{-1}$ and the mean rotational temperature is 18.1 K, both larger than 0.74 km s$^{-1}$ and 14.7 K found in cores in the nearby molecular clouds summarized by \cite{jijina1999}. The mean NH$_3$ column density is 2.3$\times$10$^{15}$ cm$^{-2}$, and the mean molecular mass is 67 M$_{\odot}$. The mean ratio between $M_{vir}$ and $M_{core}$ is 1.36, while the median ratio is 0.64. The distributions of these core properties are shown in the histograms in Figure \ref{histogram}.

The mean linewidth we found from our observations appears to be systematically smaller (by $\sim$0.5-1 km s$^{-1}$) than the linewidths measured by the single dish NH$_3$ line surveys with similar target source selection criterion \cite[e.g.][]{pillai2006,urquhart2011,wienen2012,chira2013}. However, it agrees well with the previous VLA NH$_3$ survey with a similar angular resolution but fewer samples \citep{sanchez2013}. These results can be consistent with a positive size-linewidth correlation, which will be discussed in detail in Section \ref{corre}. Given the mean gas temperature of 15-25 K, the thermal linewidth is only $\sim$0.2 km s$^{-1}$. The detected gas motion is therefore highly supersonic. We note that we resolved the molecule cores down to the scales of $\sim$0.1 pc. At this spatial scale, the observed linewidth is attributed to turbulence \citep{paredes2007} as well as infall, rotation and (proto)stellar outflows \citep{arce2007,cesaroni2007,baobab2010}.

The distribution of rotational temperature appears to be concentrated around 16 K, which is consistent with the typical temperature found in cold dense cores, e.g. $\sim$15 K in IRDCs \citep{pillai2006,chira2013}. There is another peak at 21 K, which comes from the few cores in the hot core and dispersed classes as shown in Figure \ref{histo_groups}. More samples are needed to confirm the existence of a second peak in the temperature distribution.

The detected NH$_3$ column density is generally between $5\times10^{14}$ and $5\times10^{15}$ cm$^{-2}$. Applying the abundance ratio [NH$_3$]/[H$_2$]=$3\times10^{-8}$ it leads to the H$_2$ column density of $10^{22}$-$10^{23}$ cm$^{-2}$. This translates to an extinction of $A_V$$\approx$10-100 \citep{guver2009}, in agreement with the mid-infrared extinction we saw in Figures \ref{tempfigs} \& \ref{tempfigs_more}. 

The mass distribution of the selected cores is close to lognormal with the peak around 30 $M_{\odot}$. The virial parameter $M_{vir}/M_{core}$ is wide-spread from 0.1 to 10, and the median value is 0.64 while the mean value is 1.36, both of which are close to 1.

\subsubsection{Core Properties among Different Morphological Classes}\label{3classes}

Based on the morphologies noted in Section \ref{morphology} as well as the infrared environment noted in Section \ref{infrared}, we clarified the cores into three classes: ``filaments'', taken from the filamentary sources; ``hot cores'', taken from the three confirmed hot cores; and ``dispersions'', taken from the two sources showing dispersed NH$_3$ emission around the infrared sources. The sources from which we selected the three classes are listed in Table \ref{physical}. We noted that the majority of our target sources are categorized as filaments, which leads to the largest number (80 in total) of cores in this class. There are only 6 and 5 cores identified from the hot core and dispersion classes respectively. The rest of the 174 cores cannot be classified as any of these three classes. Therefore, although we found the differences of core properties among the three classes, it requires observations with larger number of samples to justify the statistical significance.

We plotted the distributions of core properties for each group in Figure \ref{histo_groups}.  We found that the linewidth, temperature, column density, and molecular mass of cores in the hot cores are generally larger than those of cores in the dispersion class, while the filament class lies in between in these properties. The dependency of the core properties on the environment explains the bimodality in the overall temperature distribution mentioned in Section \ref{coreoverall}. The larger temperatures of the hot cores may indicate a later evolutionary stage than the other two classes \citep[cf.][Figure. 2]{hoq2013}.

The most interesting property is the virial parameter $M_{vir}/M_{core}$. It is smaller than 1 for all the cores in the hot cores class, while larger than 1 for those in the dispersion class. The virial parameter of the filaments class distributes from 0.1 to 1 almost evenly. The virial parameter appears to have a dependency among the classes, which suggests the hot cores tend to collapse due to gravity while those in the NH$_3$ dispersed sources are subject to kinematic processes, and most cores in the filamentary sources are close to the virial equilibrium.

\subsection{Correlations between Physical Properties}\label{corre}

\begin{figure}[!t]
\begin{tabular}{p{8.5cm}p{8.5cm}}
\includegraphics[width=0.47\textwidth]{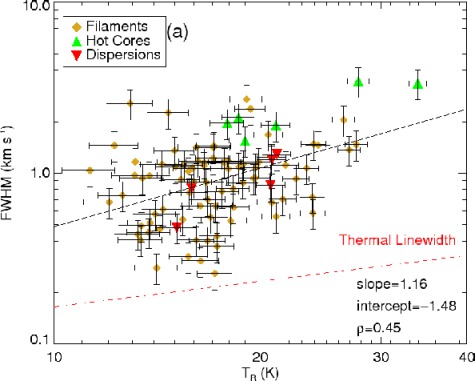} & \includegraphics[width=0.47\textwidth]{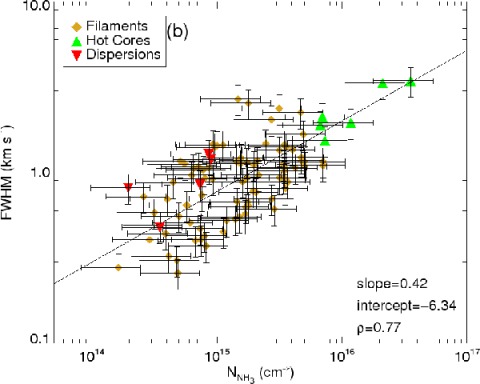} \\\includegraphics[width=0.47\textwidth]{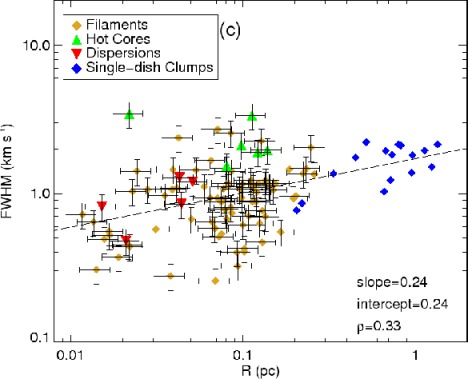} &\hspace{4pt} \includegraphics[width=0.455\textwidth]{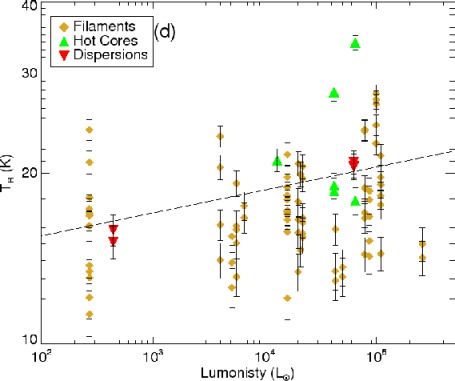}\\
\end{tabular}
\caption{Correlations between core properties. Classes of cores are shown in different symbols and colors. a). The FWHM linewidth vs. the rotational temperature. b). The FWHM linewidth vs. the NH$_3$ column density. c). The FWHM linewidth vs. the core radius. d). The cluster luminosity vs. the rotational temperature. }
\label{correlation}
\end{figure}

We plotted the temperature, column density and radius against the linewidth in Figure \ref{correlation}.  We assumed a 1-$\sigma$ error of 1 K in the temperature, 20\% in the linewidth, and 20\% and 50\% in the radius and column density, respectively. Then we performed a linear regression via Bayesian approach on the logarithmic data including all three morphological classes, to search for a power-law relation. The correlation coefficient $\rho$ of each linear regression is given in each panel. 

In Figure \ref{correlation}a, we plotted the linewidth against the rotation temperature. We found a possible correlation $\Delta v\sim T_R^{\ \ 1.16\pm0.29}$, with $\rho=0.45$. For the hot cores class, the linewidth is larger than what this relation predicts at the same temperature, while for the dispersion class it is smaller. Although the correlation is weak given the small correlation coefficient, similar correlations have been reported in large samples of single-dish observations \citep[e.g.][]{molinari1996,wu2006,urquhart2011}. We also plotted the thermal linewidth at each temperate for comparison. It is evident that the non-thermal component in the linewidth is dominant,  while the hot cores class shows the largest non-thermal contribution. 

To understand this correlation, we assumed that the temperature is determined by the radiation thus the luminosity of the protostars, while the linewidth at this scale is mainly from turbulence that maintains a virial equilibrium of the dense core. With these assumptions, we have $T_R\sim L_{\star}^{1/4}, \Delta v \sim M_{core}^{1/2}$. The luminosity of the protostars is related to their mass with $L_{\star}\sim M_{\star} ^{\alpha}$, the index $\alpha$ can be 1-3 depending on whether gas pressure or radiation pressure dominates against the gravity. Assuming a uniform star formation efficiency $M_{\star}/M_{core}$, we can relate the temperature with the linewidth by $\Delta v \sim T_R^{\gamma}$, in which $\gamma$ ranges from 0.67 to 2. Our result of $\gamma=1.16$ agrees with this simplified model. On the other hand, if the temperature is decided by turbulent heating, one can have the approximate relation $\Delta v \sim T^{1.25}$ \citep{gusten1985}, which is consistent with our result. However, in reality, both mechanisms as well as heating from accretion {\it etc.} must be taken into consideration, therefore the interpretation of this correlation requires more elaborate model, which is beyond the scope of this paper.

In Figure \ref{correlation}b, we found a relation $\Delta v\sim N_{NH_3}^{\ \ 0.42\pm0.05}$, with $\rho=0.77$. It is similar to the result of \citet{urquhart2011}, which derived a power-law index 0.24$\pm$0.07. Note that the column density $N_{NH_3}$ is intrinsically related to the linewidth as well as the rotational temperature, as indicated in Equations \ref{column} \& \ref{column_thin}, therefore unlike the other relations in this section, it is not a correlation of two independent parameters. Nonetheless, we found that the three classes all agree with the overall relation, suggesting a universal dependency of the column density on the dynamics of dense cores.

Figure \ref{correlation}c plots the radius versus the linewidth, which leads to a weak correlation $\Delta v\sim R^{\ 0.24\pm0.08}$ with $\rho=0.33$. The decrease of linewidth toward small radius may suggest the decay of turbulent motions from larger spatial scales to smaller scales. The power-law index of the correlation is smaller than those of similar studies based on single-dish observations \citep[e.g.][]{larson1981, lada1991, jijina1999, wu2006, azimlu2011}, but is consistent with the result towards nearby high-mass star forming region Orion A and B \citep{caselli1995}. For example, we compared our fitting result, i.e. the solid line in Figure \ref{correlation}c, with the Galactic NH$_3$ data taken from \citet{wu2006}, which was carried out with the 100m telescope towards similar high-mass star forming regions like ours, therefore the difference is mainly a matter of resolution. It is evident that the radius-linewidth relation we found continues to the scale of several parsec.

At last in Figure \ref{correlation}d we compared the IRAS point source luminosity $L_{IR}$ with the core temperature. Given the resolution of IRAS ($\gtrsim$1 arcmin), the luminosity should be that of the most massive stars in the cluster which dominate the energy output. Since the core temperatures were binned for each IRAS source, we performed a linear regression via minimum chi-square method in the log-log space which leads to the relation $T_R\sim L_{IR}^{\ \ 0.041\pm 0.003}$. However the probability of the correlation is $\ll$0.1 therefore the two properties are poorly correlated. For the dispersion and hot core classes, we found higher temperature towards cores associated with more luminous sources. This result is expected if they are illuminated or heated directly by the massive stars. On the contrary, for the filaments class the temperature and the luminosity are much worse correlated: the relation $T_R\sim L_{IR}^{\ \ 0.041\pm 0.003}$ is unaffected when only cores in the filaments class were taken into account, and we found low temperature cores ($T_R<$15 K) even in luminous ($L>10^4$ L$_{\odot}$) sources. This might be due to the fact that the majority of the molecule gas is still self-shielded in the filaments, therefore only the cores immediately associated with the embedded protostars which can be illuminated or be mechanically shocked are significantly heated \citep[e.g.][]{li2003,zhang2007}. Therefore, we tentatively suggested that the averaged molecular gas temperature on the 0.1 pc scale is indicative of the high-mass star forming activities for the dispersion and hot core classes. Observations of such sources with a broader range of luminosity are required to resolve the correlations between $T_R$ and $L_{IR}$.

\section{DISCUSSIONS}\label{discussions}

\subsection{The Cylinder Collapse of Filaments}\label{cylinder}

In Figures \ref{tempfigs} \& \ref{tempfigs_more} we find a handful of filamentary sources with embedded cores that are well aligned and regularly spaced. Several most convincing examples include IRAS 18372$-$0541, 18507+0121, IRAS 19074+0752, IRAS 19220+1432, IRAS 19368+2239 and IRAS 23033+5951. Such configuration has been found in other filamentary clouds \citep[e.g.][]{teixeira2006,jackson2010,zhang2009,wang2011,busquet2013}. It is consistent with what the ``sausage'' instability produces during the gravitational collapse of a cylinder \citep{chandra1953, nagasawa1987}.

According to this theory, a self-gravitational bounded cylinder, even considering the axial magnetic field, will break up into regularly spaced pieces, due to the instability growing quickly at a characteristic length interval. This interval, ${\rm \lambda_{max}}$, depends on the nature of the cylinder: if it is an incompressible fluid, ${\rm \lambda_{max}}=11R$, where $R$ is the cylinder's radius; if it is an isothermal gas cylinder, ${\rm \lambda_{max}}=22H=22v(4\pi G\rho_c)^{-1/2}$, where $H$ is the isothermal scale height and $\rho_c$ is the density at the center of the cylinder; $v$ is the isothermal sound speed $c_s$ in the case of a thermal support, or the velocity dispersion $\sigma$ in the case of a turbulent support. The mass per unit length along the cylinder also has a theoretical constraint, $(M/l)_{\rm max}$, above which the cylinder will collapse radially quickly. According to \cite{wang2011}, the critical mass per unit length is $(M/l)_{\rm max}=2v^2/G=465(\frac{v}{\rm km\ s^{-1}})^2$ M$_{\odot}$ pc$^{-1}$, in which $v$ is either $c_s$ or $\sigma$ depending the supporting mechanism. 

In IRAS 19074+0752 for instance, the filament at the south breaks into four separated cores, with an average spacing of 0.30 pc at a distance of 8.7 kpc. The total mass of this filament is 145 M$_{\odot}$ within 1.3 pc, leading to a linear mass density of 110 M$_{\odot}$ pc$^{-1}$. Note that due to the projection effect, the observed spatial interval is a lower limit of the true value while the linear mass density is an upper limit if we ignore the uncertainty in the mass.

Now for IRAS 19074+0752 we derive the theoretic predictions based on the cylinder collapse model. Before apply the models we need to determine the initial parameters of the cylinder, because once the star formation begins the environmental conditions will be changed greatly, making our estimation biased. We assume that the linewidth and temperature between the cores are closer to the initial conditions of the cylinder. For the volume density, on the one hand, the observed one is definitely larger than the initial density because of the collapse; on the other hand, it is also smaller than the density at the center. Given these two factors, we apply the average density of the filament $n$=$1.9\times10^5$ cm$^{-3}$ as a good approximation to $\rho_c$. We obtain a velocity dispersion $\sigma$ ranging from 0.35 to 0.70 km s$^{-1}$. The sound speed $c_s$ is 0.17-0.30 km s$^{-1}$ for a temperature of 8-26 K. 

First, in the case of an incompressible fluid, the radius of the filament will be $\lambda/11$. Applying $\lambda$'s observational value of 0.3 pc, we obtain a radius of 0.027 pc, much smaller than 0.15 pc, the average half-power radius of the cores. We can also estimate the strength of the magnetic field based on this radius \citep{wang2011} and the result, $\sim$0.4 mG, is smaller than typical strength in massive star formation regions \citep{girart2009}. Therefore the incompressible fluid scenario is not likely the case.

Then, for an isothermal cylinder, the spatial interval ${\rm \lambda_{max}}$ will be 0.32-0.63 pc for turbulent support, or 0.15-0.27 pc for a thermal support. The critical mass per unit length will be 57-230 M$_{\odot}$ pc$^{-1}$ for turbulent support, or 13-42 M$_{\odot}$ pc$^{-1}$ for thermal support. The results of the other five candidates are listed in Table \ref{cylinder}, which show the same trend.

The observed mass per unit length of the cylinder candidates has a consistent value of $\sim$120 M$_{\odot}$ pc$^{-1}$, outbalancing theoretical predictions when they are supported by thermal motions, but are comparable to those when supported by turbulence. Therefore thermal motions are not sufficient to support the cylinders from collapsing, while turbulent motions appear to be essential here in determining the dynamics of the filaments.
 
The observed spatial interval between cores is consistent with both the upper limit of the thermal support scenario and the lower limit of the turbulent support scenario. However, the temperature required to obtain the upper limit of $\lambda_{\rm max}$ of thermal support is probably biased by the heating of star formation. Further considering the projection effect, we suggest that the turbulent support scenario is more realistic.

Two of the cylinder candidates, IRAS 18223$-$1243 and IRAS 18507+0121, are identified as IRDCs, while the others are likely more evolved given the associations with H{\sc ii} regions or infrared emission. Compared with similar studies towards IRDCs \citep{jackson2010,wang2011,ragan2011}, the observed mass per unit length is consistent in all these sources. This suggests that the molecular gas content in the filaments is retained as they evolve from IRDCs to active star forming regions.

\subsection{Embedded Protostellar Heating Revealed by Radial Temperature Distribution}\label{hotcores}

\cite{longmore2011} used the NH$_3$ radiative transfer modeling to fit the observed NH$_3$ spectral line and derived a power-law relation $T(r)=100.3\ \left(\displaystyle\frac{r}{\rm arcsec}\right)^{-0.35}$ K towards an isolated and spatially symmetric massive star-forming core, G8.68$-$0.37. They resolved the core further into three dust continuum peaks, separated by 1-2 arcsec. 

Typical dust opacity found in star-forming regions has the dependence on the wavelength, e.g. $Q$$\sim$$\lambda^{-1.6}$ \citep{li2001}. This leads to a radial temperature distribution of $T(r)\propto r^{-0.36}$ assuming a single embedded heating source and an equilibrium between the absorption of energy from the heating source and the emission of the dust itself within a spherical shell \citep{scoville1976,garay1999}, which is consistent with the NH$_3$ gas temperature distribution in G8.68$-$0.37.

One of the hot core sources, IRAS 18360$-$0537, exhibits a temperature profile of $T(r_{prj})\propto r_{prj}^{-0.35}$. It is consistent with that in G8.68$-$0.37 and the spherical dust core model. However this relation is based on the projected radius and average temperature along the light of sight, and does not necessarily support the spherical dust core model.

We also find that two sources, IRAS 18089$-$1732 and IRAS 18182$-$1433, are best fitted with larger power-law indices and smaller intercepts, which indicates that the temperature around the protostars is lower while the decrease of the temperature with radius is less sharp, as compared to IRAS 18360$-$0537. Their luminosities are similar to that of IRAS 18360$-$0537 ($\sim$10$^4$ L$_{\odot}$). An potential interpretation is that the stronger outflows found in the two sources \citep{beuther2004b,beuther2006,qiu2012} heat the gas thus elevate the temperature $\gtrsim$0.2 pc away from the protostars \citep[cf.][]{zhang2007,wang2012}.

\section{CONCLUSIONS}\label{conclusions}

We present the NH$_3$ (J,K)=(1,1) and (2,2) inversion line observations towards a sample of 62 high-mass star formation regions with the VLA. The high angular resolution allows us to obtain spatial distributions of the rotational temperature, which reveal the thermal condition of NH$_3$ cores. Based on the morphology of the NH$_3$ emission and the infrared environment, we identify dense cores in these sources and classify them into three classes. We compare the physical properties of the cores among the classes and find possible dependence of the properties on the environment.

The main results of this paper include:

\begin{enumerate}
\item Filamentary structures with typical length of 1 pc and width of 0.1 pc are found in our sample. These filaments usually contain regularly spaced condensations along the major axis. After comparing the observed spatial intervals and mass per unit length with theoretical predictions, we found that the filaments are likely supported by turbulent motions. Particularly we found several hub-filaments systems that may form high-mass clusters via global gas accretion along the filaments.
\item The temperature peaks do not always coincide with the peaks of NH$_3$ emission, but do have an evident association with infrared sources. Therefore the temperature is related to either radiative or dynamic, or both types of feedback from protostars.
\item For the three hot core sources, we analyzed their radial temperature distribution. The projected radial distribution can be well described by a power-law relation, with the indices between $-$0.18 and $-$0.35. The relatively large index such as $-$0.18 might be caused by outflows.
\item The mean values of the NH$_3$ core properties are: the effective radius $R$$\sim$0.08 pc, the intrinsic linewidth of NH$_3$ (1,1) inversion line $\Delta v$$\sim$1.08 km s$^{-1}$, the rotational temperature $T_R$$\sim$18.1 K, the NH$_3$ column density $N({\rm NH_3})\sim2.3\times10^{15}$ cm$^{-2}$, and the molecular core mass $M_{core}$$\sim$67 M$_{\odot}$. These values are generally larger than those in low-mass star forming cores, while they are comparable with those observed by single-dish telescopes towards high-mass star forming regions. The cores found in concentrated sources tend to have larger linewidth, temperature, column density and mass than those in NH$_3$ dispersed sources, while those in filamentary sources show properties in between the other two classes. The cores in concentrated sources all have a virial parameter larger than 1, while those in dispersed sources all have a virial parameter smaller than 1, suggesting that they are in completely different dynamic states.
\item The temperature, column density and radius appear to be correlated with the linewidth, suggesting the importance of dynamics in determining the status of the dense cores. Comparisons of the correlations with those derived in studies with single-dish observations indicate they are consistent in different spatial scales.
\end{enumerate}

\acknowledgments

We are grateful to the anonymous referee for valuable suggestions. This work was partly supported by the National Basic Research Program of China (973 program) No. 2012CB821805, the National Science Foundation of China under Grant 11273015,11133001,11328301, and has made use of NASA's Astrophysics Data System, and the SIMBAD database operated at CDS, Strasbourg, France. This study used observations made with the Spitzer Space Telescope, which is operated by the Jet Propulsion Laboratory, California Institute of Technology under a contract with NASA. XL acknowledges the support of Smithsonian Predoctoral Fellowship.
    
\bibliographystyle{apj}

\begin{thebibliography}{102}
\expandafter\ifx\csname natexlab\endcsname\relax\def\natexlab#1{#1}\fi

\bibitem[{{Anderson} \& {Bania}(2009)}]{anderson2009}
{Anderson}, L.~D., \& {Bania}, T.~M. 2009, \apj, 690, 706

\bibitem[{{Arce} {et~al.}(2007){Arce}, {Shepherd}, {Gueth}, {Lee}, {Bachiller},
  {Rosen}, \& {Beuther}}]{arce2007}
{Arce}, H.~G., {Shepherd}, D., {Gueth}, F., {et~al.} 2007, Protostars and
  Planets V, 245

\bibitem[{{Argon} {et~al.}(2000){Argon}, {Reid}, \& {Menten}}]{argon2000}
{Argon}, A.~L., {Reid}, M.~J., \& {Menten}, K.~M. 2000, \apjs, 129, 159

\bibitem[{{Azimlu} \& {Fich}(2011)}]{azimlu2011}
{Azimlu}, M., \& {Fich}, M. 2011, \aj, 141, 123

\bibitem[{{Ballesteros-Paredes} {et~al.}(2007){Ballesteros-Paredes}, {Klessen},
  {Mac Low}, \& {Vazquez-Semadeni}}]{paredes2007}
{Ballesteros-Paredes}, J., {Klessen}, R.~S., {Mac Low}, M.-M., \&
  {Vazquez-Semadeni}, E. 2007, Protostars and Planets V, 63

\bibitem[{{Barranco} \& {Goodman}(1998)}]{barranco1998}
{Barranco}, J.~A., \& {Goodman}, A.~A. 1998, \apj, 504, 207

\bibitem[{{Bartkiewicz} {et~al.}(2009){Bartkiewicz}, {Szymczak}, {van
  Langevelde}, {Richards}, \& {Pihlstr{\"o}m}}]{bartkiewicz2009}
{Bartkiewicz}, A., {Szymczak}, M., {van Langevelde}, H.~J., {Richards},
  A.~M.~S., \& {Pihlstr{\"o}m}, Y.~M. 2009, \aap, 502, 155

\bibitem[{{Bergin} \& {Langer}(1997)}]{bergin1997}
{Bergin}, E.~A., \& {Langer}, W.~D. 1997, \apj, 486, 316

\bibitem[{{Beuther} {et~al.}(2007){Beuther}, {Leurini}, {Schilke}, {Wyrowski},
  {Menten}, \& {Zhang}}]{beuther2007}
{Beuther}, H., {Leurini}, S., {Schilke}, P., {et~al.} 2007, \aap, 466, 1065

\bibitem[{{Beuther} {et~al.}(2002{\natexlab{a}}){Beuther}, {Schilke}, {Menten},
  {Motte}, {Sridharan}, \& {Wyrowski}}]{beuther2002cs}
{Beuther}, H., {Schilke}, P., {Menten}, K.~M., {et~al.} 2002{\natexlab{a}},
  \apj, 566, 945

\bibitem[{{Beuther} {et~al.}(2002{\natexlab{b}}){Beuther}, {Walsh}, {Schilke},
  {Sridharan}, {Menten}, \& {Wyrowski}}]{beuther2002maser}
{Beuther}, H., {Walsh}, A., {Schilke}, P., {et~al.} 2002{\natexlab{b}}, \aap,
  390, 289

\bibitem[{{Beuther} {et~al.}(2006){Beuther}, {Zhang}, {Sridharan}, {Lee}, \&
  {Zapata}}]{beuther2006}
{Beuther}, H., {Zhang}, Q., {Sridharan}, T.~K., {Lee}, C.-F., \& {Zapata},
  L.~A. 2006, \aap, 454, 221

\bibitem[{{Beuther} {et~al.}(2004{\natexlab{a}}){Beuther}, {Zhang}, {Hunter},
  {Sridharan}, {Zhao}, {Sollins}, {Ho}, {Liu}, {Ohashi}, {Su}, \&
  {Lim}}]{beuther2004a}
{Beuther}, H., {Zhang}, Q., {Hunter}, T.~R., {et~al.} 2004{\natexlab{a}},
  \apjl, 616, L19

\bibitem[{{Beuther} {et~al.}(2004{\natexlab{b}}){Beuther}, {Hunter}, {Zhang},
  {Sridharan}, {Zhao}, {Sollins}, {Ho}, {Ohashi}, {Su}, {Lim}, \&
  {Liu}}]{beuther2004b}
{Beuther}, H., {Hunter}, T.~R., {Zhang}, Q., {et~al.} 2004{\natexlab{b}},
  \apjl, 616, L23

\bibitem[{{Busquet} {et~al.}(2013){Busquet}, {Zhang}, {Palau}, {Liu},
  {S{\'a}nchez-Monge}, {Estalella}, {Ho}, {de Gregorio-Monsalvo}, {Pillai},
  {Wyrowski}, {Girart}, {Santos}, \& {Franco}}]{busquet2013}
{Busquet}, G., {Zhang}, Q., {Palau}, A., {et~al.} 2013, \apjl, 764, L26

\bibitem[{{Caselli} \& {Myers}(1995)}]{caselli1995}
{Caselli}, P., \& {Myers}, P.~C. 1995, \apj, 446, 665

\bibitem[{{Caswell}(1998)}]{caswell1998}
{Caswell}, J.~L. 1998, \mnras, 297, 215

\bibitem[{{Cesaroni} {et~al.}(2007){Cesaroni}, {Galli}, {Lodato}, {Walmsley},
  \& {Zhang}}]{cesaroni2007}
{Cesaroni}, R., {Galli}, D., {Lodato}, G., {Walmsley}, C.~M., \& {Zhang}, Q.
  2007, Protostars and Planets V, 197

\bibitem[{{Chandrasekhar} \& {Fermi}(1953)}]{chandra1953}
{Chandrasekhar}, S., \& {Fermi}, E. 1953, \apj, 118, 116

\bibitem[{{Chira} {et~al.}(2013){Chira}, {Beuther}, {Linz}, {Schuller},
  {Walmsley}, {Menten}, \& {Bronfman}}]{chira2013}
{Chira}, R.-A., {Beuther}, H., {Linz}, H., {et~al.} 2013, \aap, 552, A40

\bibitem[{{Churchwell} {et~al.}(1990){Churchwell}, {Walmsley}, \&
  {Cesaroni}}]{churchwell1990}
{Churchwell}, E., {Walmsley}, C.~M., \& {Cesaroni}, R. 1990, \aaps, 83, 119

\bibitem[{{Churchwell} {et~al.}(2006){Churchwell}, {Povich}, {Allen}, {Taylor},
  {Meade}, {Babler}, {Indebetouw}, {Watson}, {Whitney}, {Wolfire}, {Bania},
  {Benjamin}, {Clemens}, {Cohen}, {Cyganowski}, {Jackson}, {Kobulnicky},
  {Mathis}, {Mercer}, {Stolovy}, {Uzpen}, {Watson}, \&
  {Wolff}}]{churchwell2006}
{Churchwell}, E., {Povich}, M.~S., {Allen}, D., {et~al.} 2006, \apj, 649, 759

\bibitem[{{Csengeri} {et~al.}(2011){Csengeri}, {Bontemps}, {Schneider},
  {Motte}, \& {Dib}}]{csengeri2011}
{Csengeri}, T., {Bontemps}, S., {Schneider}, N., {Motte}, F., \& {Dib}, S.
  2011, \aap, 527, A135

\bibitem[{{Edris} {et~al.}(2011){Edris}, {Fuller}, {Etoka}, \&
  {Cohen}}]{edris2011}
{Edris}, K.~A., {Fuller}, G.~A., {Etoka}, S., \& {Cohen}, R.~J. 2011, ArXiv
  e-prints

\bibitem[{{Fazal} {et~al.}(2008){Fazal}, {Sridharan}, {Qiu}, {Robitaille},
  {Whitney}, \& {Zhang}}]{fazal2008}
{Fazal}, F.~M., {Sridharan}, T.~K., {Qiu}, K., {et~al.} 2008, \apjl, 688, L41

\bibitem[{{Fazio} {et~al.}(2004){Fazio}, {Hora}, {Allen}, {Ashby}, {Barmby},
  {Deutsch}, {Huang}, {Kleiner}, {Marengo}, {Megeath}, {Melnick}, {Pahre},
  {Patten}, {Polizotti}, {Smith}, {Taylor}, {Wang}, {Willner}, {Hoffmann},
  {Pipher}, {Forrest}, {McMurty}, {McCreight}, {McKelvey}, {McMurray}, {Koch},
  {Moseley}, {Arendt}, {Mentzell}, {Marx}, {Losch}, {Mayman}, {Eichhorn},
  {Krebs}, {Jhabvala}, {Gezari}, {Fixsen}, {Flores}, {Shakoorzadeh}, {Jungo},
  {Hakun}, {Workman}, {Karpati}, {Kichak}, {Whitley}, {Mann}, {Tollestrup},
  {Eisenhardt}, {Stern}, {Gorjian}, {Bhattacharya}, {Carey}, {Nelson},
  {Glaccum}, {Lacy}, {Lowrance}, {Laine}, {Reach}, {Stauffer}, {Surace},
  {Wilson}, {Wright}, {Hoffman}, {Domingo}, \& {Cohen}}]{fazio2004}
{Fazio}, G.~G., {Hora}, J.~L., {Allen}, L.~E., {et~al.} 2004, \apjs, 154, 10

\bibitem[{{Fish} {et~al.}(2003){Fish}, {Reid}, {Wilner}, \&
  {Churchwell}}]{fish2003}
{Fish}, V.~L., {Reid}, M.~J., {Wilner}, D.~J., \& {Churchwell}, E. 2003, \apj,
  587, 701

\bibitem[{{Fontani} {et~al.}(2010){Fontani}, {Cesaroni}, \&
  {Furuya}}]{fontani2010}
{Fontani}, F., {Cesaroni}, R., \& {Furuya}, R.~S. 2010, \aap, 517, A56

\bibitem[{{Forster} \& {Caswell}(1989)}]{forster1989}
{Forster}, J.~R., \& {Caswell}, J.~L. 1989, \aap, 213, 339

\bibitem[{{Galv{\'a}n-Madrid} {et~al.}(2010){Galv{\'a}n-Madrid}, {Zhang},
  {Keto}, {Ho}, {Zapata}, {Rodr{\'{\i}}guez}, {Pineda}, \&
  {V{\'a}zquez-Semadeni}}]{galvan2010}
{Galv{\'a}n-Madrid}, R., {Zhang}, Q., {Keto}, E., {et~al.} 2010, \apj, 725, 17

\bibitem[{{Galv{\'a}n-Madrid} {et~al.}(2013){Galv{\'a}n-Madrid}, {Liu},
  {Zhang}, {Pineda}, {Peng}, {Zhang}, {Keto}, {Ho}, {Rodriguez}, {Zapata},
  {Peters}, {De Pree}, \& {.}}]{galvan2013}
{Galv{\'a}n-Madrid}, R., {Liu}, H.~B., {Zhang}, Z.-Y., {et~al.} 2013, ArXiv
  e-prints

\bibitem[{{Garay} \& {Lizano}(1999)}]{garay1999}
{Garay}, G., \& {Lizano}, S. 1999, \pasp, 111, 1049

\bibitem[{{Girart} {et~al.}(2009){Girart}, {Beltr{\'a}n}, {Zhang}, {Rao}, \&
  {Estalella}}]{girart2009}
{Girart}, J.~M., {Beltr{\'a}n}, M.~T., {Zhang}, Q., {Rao}, R., \& {Estalella},
  R. 2009, Science, 324, 1408

\bibitem[{{Guesten} {et~al.}(1985){Guesten}, {Walmsley}, {Ungerechts}, \&
  {Churchwell}}]{gusten1985}
{Guesten}, R., {Walmsley}, C.~M., {Ungerechts}, H., \& {Churchwell}, E. 1985,
  \aap, 142, 381

\bibitem[{{G{\"u}ver} \& {{\"O}zel}(2009)}]{guver2009}
{G{\"u}ver}, T., \& {{\"O}zel}, F. 2009, \mnras, 400, 2050

\bibitem[{{Harju} {et~al.}(1998){Harju}, {Lehtinen}, {Booth}, \&
  {Zinchenko}}]{harju1998}
{Harju}, J., {Lehtinen}, K., {Booth}, R.~S., \& {Zinchenko}, I. 1998, \aaps,
  132, 211

\bibitem[{{Harju} {et~al.}(1993){Harju}, {Walmsley}, \&
  {Wouterloot}}]{harju1993}
{Harju}, J., {Walmsley}, C.~M., \& {Wouterloot}, J.~G.~A. 1993, \aaps, 98, 51

\bibitem[{{Ho} \& {Townes}(1983)}]{Ho1983}
{Ho}, P.~T.~P., \& {Townes}, C.~H. 1983, \araa, 21, 239

\bibitem[{{Hoq} {et~al.}(2013){Hoq}, {Jackson}, {Foster}, {Sanhueza}, {Guzman},
  {Whitaker}, {Claysmith}, {Rathborne}, {Vasyunina}, \& {Vasyunin}}]{hoq2013}
{Hoq}, S., {Jackson}, J.~M., {Foster}, J.~B., {et~al.} 2013, ArXiv e-prints

\bibitem[{{Jackson} {et~al.}(2010){Jackson}, {Finn}, {Chambers}, {Rathborne},
  \& {Simon}}]{jackson2010}
{Jackson}, J.~M., {Finn}, S.~C., {Chambers}, E.~T., {Rathborne}, J.~M., \&
  {Simon}, R. 2010, \apjl, 719, L185

\bibitem[{{Jijina} {et~al.}(1999){Jijina}, {Myers}, \& {Adams}}]{jijina1999}
{Jijina}, J., {Myers}, P.~C., \& {Adams}, F.~C. 1999, \apjs, 125, 161

\bibitem[{{Keto} \& {Zhang}(2010)}]{keto2010}
{Keto}, E., \& {Zhang}, Q. 2010, \mnras, 406, 102

\bibitem[{{Krumholz} {et~al.}(2014){Krumholz}, {Bate}, {Arce}, {Dale},
  {Gutermuth}, {Klein}, {Li}, {Nakamura}, \& {Zhang}}]{krumholz2014}
{Krumholz}, M.~R., {Bate}, M.~R., {Arce}, H.~G., {et~al.} 2014, ArXiv e-prints

\bibitem[{{Kurayama} {et~al.}(2011){Kurayama}, {Nakagawa}, {Sawada-Satoh},
  {Sato}, {Honma}, {Sunada}, {Hirota}, \& {Imai}}]{kurayama2011}
{Kurayama}, T., {Nakagawa}, A., {Sawada-Satoh}, S., {et~al.} 2011, \pasj, 63,
  513

\bibitem[{{Lada} {et~al.}(1991){Lada}, {Bally}, \& {Stark}}]{lada1991}
{Lada}, E.~A., {Bally}, J., \& {Stark}, A.~A. 1991, \apj, 368, 432

\bibitem[{{Larson}(1981)}]{larson1981}
{Larson}, R.~B. 1981, \mnras, 194, 809

\bibitem[{{Li} \& {Draine}(2001)}]{li2001}
{Li}, A., \& {Draine}, B.~T. 2001, \apj, 554, 778

\bibitem[{{Li} {et~al.}(2003){Li}, {Goldsmith}, \& {Menten}}]{li2003}
{Li}, D., {Goldsmith}, P.~F., \& {Menten}, K. 2003, \apj, 587, 262

\bibitem[{{Li} {et~al.}(2012){Li}, {Wang}, {Gu}, {Zhang}, \&
  {Zheng}}]{lijuan2012}
{Li}, J., {Wang}, J., {Gu}, Q., {Zhang}, Z.-y., \& {Zheng}, X. 2012, \apj, 745,
  47

\bibitem[{{Liu} {et~al.}(2010){Liu}, {Ho}, \& {Zhang}}]{baobab2010}
{Liu}, H.~B., {Ho}, P.~T.~P., \& {Zhang}, Q. 2010, \apj, 725, 2190

\bibitem[{{Liu} {et~al.}(2012{\natexlab{a}}){Liu}, {Jim{\'e}nez-Serra}, {Ho},
  {Chen}, {Zhang}, \& {Li}}]{baobab2012a}
{Liu}, H.~B., {Jim{\'e}nez-Serra}, I., {Ho}, P.~T.~P., {et~al.}
  2012{\natexlab{a}}, \apj, 756, 10

\bibitem[{{Liu} {et~al.}(2012{\natexlab{b}}){Liu}, {Quintana-Lacaci}, {Wang},
  {Ho}, {Li}, {Zhang}, \& {Zhang}}]{baobab2012b}
{Liu}, H.~B., {Quintana-Lacaci}, G., {Wang}, K., {et~al.} 2012{\natexlab{b}},
  \apj, 745, 61

\bibitem[{{Liu} {et~al.}(2011){Liu}, {Zhang}, \& {Ho}}]{baobab2011}
{Liu}, H.~B., {Zhang}, Q., \& {Ho}, P.~T.~P. 2011, \apj, 729, 100

\bibitem[{{Longmore} {et~al.}(2007){Longmore}, {Burton}, {Barnes}, {Wong},
  {Purcell}, \& {Ott}}]{longmore2007}
{Longmore}, S.~N., {Burton}, M.~G., {Barnes}, P.~J., {et~al.} 2007, \mnras,
  379, 535

\bibitem[{{Longmore} {et~al.}(2011){Longmore}, {Pillai}, {Keto}, {Zhang}, \&
  {Qiu}}]{longmore2011}
{Longmore}, S.~N., {Pillai}, T., {Keto}, E., {Zhang}, Q., \& {Qiu}, K. 2011,
  \apj, 726, 97

\bibitem[{{MacLaren} {et~al.}(1988){MacLaren}, {Richardson}, \&
  {Wolfendale}}]{maclaren1988}
{MacLaren}, I., {Richardson}, K.~M., \& {Wolfendale}, A.~W. 1988, \apj, 333,
  821

\bibitem[{{Mangum} {et~al.}(1992){Mangum}, {Wootten}, \& {Mundy}}]{mangum1992}
{Mangum}, J.~G., {Wootten}, A., \& {Mundy}, L.~G. 1992, \apj, 388, 467

\bibitem[{{McKee} \& {Tan}(2003)}]{mckee2003}
{McKee}, C.~F., \& {Tan}, J.~C. 2003, \apj, 585, 850

\bibitem[{{Menten} \& {van der Tak}(2004)}]{menten2004}
{Menten}, K.~M., \& {van der Tak}, F.~F.~S. 2004, \aap, 414, 289

\bibitem[{{Minier} {et~al.}(2000){Minier}, {Booth}, \& {Conway}}]{minier2000}
{Minier}, V., {Booth}, R.~S., \& {Conway}, J.~E. 2000, \aap, 362, 1093

\bibitem[{{Molinari} {et~al.}(1996){Molinari}, {Brand}, {Cesaroni}, \&
  {Palla}}]{molinari1996}
{Molinari}, S., {Brand}, J., {Cesaroni}, R., \& {Palla}, F. 1996, \aap, 308,
  573

\bibitem[{{Myers}(2009)}]{myers2009}
{Myers}, P.~C. 2009, \apj, 700, 1609

\bibitem[{{Myers}(2011)}]{myers2011}
---. 2011, \apj, 735, 82

\bibitem[{{Nagasawa}(1987)}]{nagasawa1987}
{Nagasawa}, M. 1987, Progress of Theoretical Physics, 77, 635

\bibitem[{{Niinuma} {et~al.}(2011){Niinuma}, {Nagayama}, {Hirota}, {Honma},
  {Motogi}, {Nakagawa}, {Kurayama}, {Kan-Ya}, {Kawaguchi}, {Kobayashi}, \&
  {Ueno}}]{niinuma2011}
{Niinuma}, K., {Nagayama}, T., {Hirota}, T., {et~al.} 2011, \pasj, 63, 9

\bibitem[{{Palla} {et~al.}(1991){Palla}, {Brand}, {Comoretto}, {Felli}, \&
  {Cesaroni}}]{palla1991}
{Palla}, F., {Brand}, J., {Comoretto}, G., {Felli}, M., \& {Cesaroni}, R. 1991,
  \aap, 246, 249

\bibitem[{{Peeters}(2011)}]{peeters2011}
{Peeters}, E. 2011, in IAU Symposium, Vol. 280, IAU Symposium, 149--161

\bibitem[{{Peretto} {et~al.}(2013){Peretto}, {Fuller}, {Duarte-Cabral},
  {Avison}, {Hennebelle}, {Pineda}, {Andr{\'e}}, {Bontemps}, {Motte},
  {Schneider}, \& {Molinari}}]{peretto2013}
{Peretto}, N., {Fuller}, G.~A., {Duarte-Cabral}, A., {et~al.} 2013, \aap, 555,
  A112

\bibitem[{{Pestalozzi} {et~al.}(2005){Pestalozzi}, {Minier}, \&
  {Booth}}]{pestalozzi2005}
{Pestalozzi}, M.~R., {Minier}, V., \& {Booth}, R.~S. 2005, \aap, 432, 737

\bibitem[{{Pillai} {et~al.}(2006){Pillai}, {Wyrowski}, {Carey}, \&
  {Menten}}]{pillai2006}
{Pillai}, T., {Wyrowski}, F., {Carey}, S.~J., \& {Menten}, K.~M. 2006, \aap,
  450, 569

\bibitem[{{Price} {et~al.}(2001){Price}, {Egan}, {Carey}, {Mizuno}, \&
  {Kuchar}}]{price2001}
{Price}, S.~D., {Egan}, M.~P., {Carey}, S.~J., {Mizuno}, D.~R., \& {Kuchar},
  T.~A. 2001, \aj, 121, 2819

\bibitem[{{Qiu} {et~al.}(2012){Qiu}, {Zhang}, {Beuther}, \&
  {Fallscheer}}]{qiu2012}
{Qiu}, K., {Zhang}, Q., {Beuther}, H., \& {Fallscheer}, C. 2012, \apj, 756, 170

\bibitem[{{Ragan} {et~al.}(2011){Ragan}, {Bergin}, \& {Wilner}}]{ragan2011}
{Ragan}, S.~E., {Bergin}, E.~A., \& {Wilner}, D. 2011, \apj, 736, 163

\bibitem[{{Reid} {et~al.}(2009){Reid}, {Menten}, {Zheng}, {Brunthaler},
  {Moscadelli}, {Xu}, {Zhang}, {Sato}, {Honma}, {Hirota}, {Hachisuka}, {Choi},
  {Moellenbrock}, \& {Bartkiewicz}}]{Reid2009}
{Reid}, M.~J., {Menten}, K.~M., {Zheng}, X.~W., {et~al.} 2009, \apj, 700, 137

\bibitem[{{Rohlfs} \& {Wilson}(2004)}]{rohlfs2004}
{Rohlfs}, K., \& {Wilson}, T.~L. 2004, {Tools of radio astronomy, 4th rev.~and
  enl.~ed., Berlin: Springer}

\bibitem[{{S{\'a}nchez-Monge} {et~al.}(2013){S{\'a}nchez-Monge}, {Palau},
  {Fontani}, {Busquet}, {Ju{\'a}rez}, {Estalella}, {Tan}, {Sep{\'u}lveda},
  {Ho}, {Zhang}, \& {Kurtz}}]{sanchez2013}
{S{\'a}nchez-Monge}, {\'A}., {Palau}, A., {Fontani}, F., {et~al.} 2013, \mnras

\bibitem[{{Sanders} \& {Mirabel}(1996)}]{sanders1996}
{Sanders}, D.~B., \& {Mirabel}, I.~F. 1996, \araa, 34, 749

\bibitem[{{Sato} {et~al.}(2010){Sato}, {Hirota}, {Reid}, {Honma}, {Kobayashi},
  {Iwadate}, {Miyaji}, \& {Shibata}}]{sato2010}
{Sato}, M., {Hirota}, T., {Reid}, M.~J., {et~al.} 2010, \pasj, 62, 287

\bibitem[{{Schutte} {et~al.}(1993){Schutte}, {van der Walt}, {Gaylard}, \&
  {MacLeod}}]{schutte1993}
{Schutte}, A.~J., {van der Walt}, D.~J., {Gaylard}, M.~J., \& {MacLeod}, G.~C.
  1993, \mnras, 261, 783

\bibitem[{{Scoville} \& {Kwan}(1976)}]{scoville1976}
{Scoville}, N.~Z., \& {Kwan}, J. 1976, \apj, 206, 718

\bibitem[{{Sridharan} {et~al.}(2002){Sridharan}, {Beuther}, {Schilke},
  {Menten}, \& {Wyrowski}}]{sridharan2002}
{Sridharan}, T.~K., {Beuther}, H., {Schilke}, P., {Menten}, K.~M., \&
  {Wyrowski}, F. 2002, \apj, 566, 931

\bibitem[{{Szymczak} {et~al.}(2000){Szymczak}, {Hrynek}, \&
  {Kus}}]{szymczak2000}
{Szymczak}, M., {Hrynek}, G., \& {Kus}, A.~J. 2000, \aaps, 143, 269

\bibitem[{{Teixeira} {et~al.}(2006){Teixeira}, {Lada}, {Young}, {Marengo},
  {Muench}, {Muzerolle}, {Siegler}, {Rieke}, {Hartmann}, {Megeath}, \&
  {Fazio}}]{teixeira2006}
{Teixeira}, P.~S., {Lada}, C.~J., {Young}, E.~T., {et~al.} 2006, \apjl, 636,
  L45

\bibitem[{{Ulrich}(1976)}]{ulrich1976}
{Ulrich}, R.~K. 1976, \apj, 210, 377

\bibitem[{{Urquhart} {et~al.}(2011){Urquhart}, {Morgan}, {Figura}, {Moore},
  {Lumsden}, {Hoare}, {Oudmaijer}, {Mottram}, {Davies}, \&
  {Dunham}}]{urquhart2011}
{Urquhart}, J.~S., {Morgan}, L.~K., {Figura}, C.~C., {et~al.} 2011, \mnras,
  418, 1689

\bibitem[{{Val'tts} {et~al.}(2000){Val'tts}, {Ellingsen}, {Slysh}, {Kalenskii},
  {Otrupcek}, \& {Larionov}}]{valtts2000}
{Val'tts}, I.~E., {Ellingsen}, S.~P., {Slysh}, V.~I., {et~al.} 2000, \mnras,
  317, 315

\bibitem[{{Walmsley} \& {Ungerechts}(1983)}]{walmsley1983}
{Walmsley}, C.~M., \& {Ungerechts}, H. 1983, \aap, 122, 164

\bibitem[{{Walsh} {et~al.}(1998){Walsh}, {Burton}, {Hyland}, \&
  {Robinson}}]{walsh1998}
{Walsh}, A.~J., {Burton}, M.~G., {Hyland}, A.~R., \& {Robinson}, G. 1998,
  \mnras, 301, 640

\bibitem[{{Wang} {et~al.}(2012){Wang}, {Zhang}, {Wu}, {Li}, \&
  {Zhang}}]{wang2012}
{Wang}, K., {Zhang}, Q., {Wu}, Y., {Li}, H.-b., \& {Zhang}, H. 2012, \apjl,
  745, L30

\bibitem[{{Wang} {et~al.}(2011){Wang}, {Zhang}, {Wu}, \& {Zhang}}]{wang2011}
{Wang}, K., {Zhang}, Q., {Wu}, Y., \& {Zhang}, H. 2011, \apj, 735, 64

\bibitem[{{Wang} {et~al.}(2007){Wang}, {Wu}, {Zhang}, {Mao}, \&
  {Miller}}]{wang2007}
{Wang}, Y., {Wu}, Y., {Zhang}, Q., {Mao}, R.-Q., \& {Miller}, M. 2007, \aap,
  461, 197

\bibitem[{{Werner} {et~al.}(2004){Werner}, {Roellig}, {Low}, {Rieke}, {Rieke},
  {Hoffmann}, {Young}, {Houck}, {Brandl}, {Fazio}, {Hora}, {Gehrz}, {Helou},
  {Soifer}, {Stauffer}, {Keene}, {Eisenhardt}, {Gallagher}, {Gautier}, {Irace},
  {Lawrence}, {Simmons}, {Van Cleve}, {Jura}, {Wright}, \&
  {Cruikshank}}]{werner2004}
{Werner}, M.~W., {Roellig}, T.~L., {Low}, F.~J., {et~al.} 2004, \apjs, 154, 1

\bibitem[{{Wienen} {et~al.}(2012){Wienen}, {Wyrowski}, {Schuller}, {Menten},
  {Walmsley}, {Bronfman}, \& {Motte}}]{wienen2012}
{Wienen}, M., {Wyrowski}, F., {Schuller}, F., {et~al.} 2012, \aap, 544, A146

\bibitem[{{Wilking} {et~al.}(1989){Wilking}, {Blackwell}, {Mundy}, \&
  {Howe}}]{wilking1989}
{Wilking}, B.~A., {Blackwell}, J.~H., {Mundy}, L.~G., \& {Howe}, J.~E. 1989,
  \apj, 345, 257

\bibitem[{{Williams} {et~al.}(1994){Williams}, {de Geus}, \&
  {Blitz}}]{williams1994}
{Williams}, J.~P., {de Geus}, E.~J., \& {Blitz}, L. 1994, \apj, 428, 693

\bibitem[{{Wu} {et~al.}(2006){Wu}, {Zhang}, {Yu}, {Miller}, {Mao}, {Sun}, \&
  {Wang}}]{wu2006}
{Wu}, Y., {Zhang}, Q., {Yu}, W., {et~al.} 2006, \aap, 450, 607

\bibitem[{{Zhang} {et~al.}(1998){Zhang}, {Hunter}, \& {Sridharan}}]{zhang1998}
{Zhang}, Q., {Hunter}, T.~R., \& {Sridharan}, T.~K. 1998, \apjl, 505, L151

\bibitem[{{Zhang} {et~al.}(1999){Zhang}, {Hunter}, {Sridharan}, \&
  {Cesaroni}}]{zhang1999}
{Zhang}, Q., {Hunter}, T.~R., {Sridharan}, T.~K., \& {Cesaroni}, R. 1999,
  \apjl, 527, L117

\bibitem[{{Zhang} {et~al.}(2002){Zhang}, {Hunter}, {Sridharan}, \&
  {Ho}}]{zhang2002}
{Zhang}, Q., {Hunter}, T.~R., {Sridharan}, T.~K., \& {Ho}, P.~T.~P. 2002, \apj,
  566, 982

\bibitem[{{Zhang} {et~al.}(2007){Zhang}, {Sridharan}, {Hunter}, {Chen},
  {Beuther}, \& {Wyrowski}}]{zhang2007}
{Zhang}, Q., {Sridharan}, T.~K., {Hunter}, T.~R., {et~al.} 2007, \aap, 470, 269

\bibitem[{{Zhang} {et~al.}(2009){Zhang}, {Wang}, {Pillai}, \&
  {Rathborne}}]{zhang2009}
{Zhang}, Q., {Wang}, Y., {Pillai}, T., \& {Rathborne}, J. 2009, \apj, 696, 268

\bibitem[{{Zinchenko} {et~al.}(1997){Zinchenko}, {Henning}, \&
  {Schreyer}}]{zinchenko1997}
{Zinchenko}, I., {Henning}, T., \& {Schreyer}, K. 1997, \aaps, 124, 385

\end{thebibliography}

\clearpage

\clearpage

\clearpage

\clearpage

\clearpage

\clearpage

\begin{deluxetable}{cccccccc}
\tabletypesize{\scriptsize}
\tablecaption{Summary of the VLA observations. \label{obssummary}}
\tablewidth{0pt}
\tablehead{
\multirow{2}{*}{\it Index} & \multirow{2}{*}{Project} & \multirow{2}{*}{Date} & \multirow{2}{*}{Config.} & $\it uv$ Distance & \multirow{2}{*}{Number} & \multirow{2}{*}{Phase Calibrators} & RMS Level \\
 & & & & (k$\lambda$) & & & (mJy beam$^{-1}$ per channel)
}
\startdata
1 & AZ0103 &1997 Dec 13 &  D & 2 - 77 & 1 & 1923+210 & 8.5 \\
2 & AS0708 &2001 Jul 23 &  C & 2 - 268 & 1 & 1829$-$106 & 7.0 \\
3 & AS0747 & 2003 Jan 16 & DnC & 2 - 167 & 5 & 1829$-$106, 1849+005 & 2.2 \\
4 & AC0733 & 2004 Jun 14/15 & DnC & 2 - 167 & 42 & 0530+135, 1743$-$038, & 2.4 \\
& & & & & & 1851+005, 1925+211 & \\
5 & AC0733 & 2004 Jun 21 & D & 2 - 81 & 12 & 1851+005, 2322+509 & 2.2 \\
6 &AW0666 & 2005 Nov 12 & D & 2 - 81 & 1 & 1832$-$105 & 2.2
\enddata
\end{deluxetable}

\clearpage

\begin{deluxetable}{cccccccrcc}
\tabletypesize{\scriptsize}
\tablecaption{Observational properties of the sources.\label{basicproperties}}
\tablewidth{0pt}
\tablehead{
\multirow{2}{*}{Source Name} & {\it Project}\tablenotemark{a} & RA\tablenotemark{b} &Dec\tablenotemark{b} & {\it l}\tablenotemark{b} &{\it b}\tablenotemark{b} &
\multirow{2}{*}{NH$_3$ Detection} & V$_{\rm LSR}$ & Distance\tablenotemark{c} & {\it L$_{IR}$} \\
 & {\it Index} & (J2000) & (J2000) & (deg) & (deg) &
 & (km/s) & (kpc) & ({\it L$_{\odot}$})
}
\startdata
IRAS 05358+3543 & 4 & 05:39:12.60 & +35:45:52.30 & 173.481 & 2.445 & (1,1),(2,2) & $-$15.1 & 1.8$^{1}$ & 6.6$\times$10$^3$   \\
IRAS 05490+2658 & 4 & 05:52:12.30 & +26:59:37.40 & 182.414 & 0.245 & (1,1) & 0.8 & 2.1$^{1}$ & 4.4$\times$10$^3$ \\
IRAS 05553+1631 & 4 & 05:58:13.90 & +16:32:00.00 & 192.161 & $-$3.815 & \nodata &\nodata &\nodata &\nodata \\
IRAS 06061+2151 & 4 & 06:09:07.80 & +21:50:39.00 & 188.796 & 1.033 & (1,1),(2,2) & $-$1.2 & 2.0$^{2}$ & 9.9$\times$10$^3$   \\
IRAS 06103+1523 & 4 & 06:13:15.10 & +15:22:36.50 & 194.934 & $-$1.227 & (1,1),(2,2) & 15.6 & 2.7$^{3}$  & 5.0$\times$10$^3$   \\
IRAS 06382+0939 & 4 & 06:41:02.70 & +09:36:10.40 & 203.205 & 2.080 & (1,1) & 5.8 & 0.8$^{4}$ & 4.4$\times$10$^2$ \\
IRAS 06501+0143 & 4 & 06:52:45.50 & +01:40:31.90 & 211.591 & 1.058 & \nodata &\nodata &\nodata &\nodata \\
IRAS 18048$-$2019 & 4 & 18:07:51.60 & $-$20:18:36.10 & 9.994 & $-$0.030 & (1,1),(2,2) & 48.5 & 4.7 & 1.9$\times$10$^4$   \\
IRAS 18089$-$1732 & 3 & 18:11:51.27 & $-$17:31:28.55 & 12.889 & 0.490 & (1,1),(2,2) & 32.0 & 3.3 & 4.2$\times$10$^4$ \\
IRAS 18090$-$1832 & 4 & 18:12:01.87 & $-$18:31:54.80 & 12.026 & $-$0.031 & (1,1),(2,2) & 110.4 & 6.3 & 4.7$\times$10$^4$   \\
W33A & 4 & 18:14:39.50 & $-$17:51:59.80 & 12.909 & $-$0.260 & (1,1),(2,2) & 38.4 & 3.7 & 1.1$\times$10$^5$   \\
IRAS 18151$-$1208 & 3 & 18:17:58.02 & $-$12:07:21.62 & 18.342 & 1.769 & (1,1),(2,2) & 32.8 & 2.8$^{5}$ & 2.2$\times$10$^4$   \\
IRAS 18159$-$1550 & 4 & 18:18:48.13 & $-$15:49:05.70 & 15.182 & $-$0.159 & (1,1),(2,2) & 59.9 & 4.5 & 2.2$\times$10$^4$   \\
IRAS 18159$-$1648 & 4 & 18:18:53.50 & $-$16:47:38.80 & 14.332 & $-$0.639 & (1,1),(2,2) & 22.1 & 1.1$^{6}$ & 4.0$\times$10$^3$   \\
IRAS 18182$-$1433 & 3 & 18:21:07.66 & $-$14:31:52.95 & 16.581 & $-$0.046 & (1,1),(2,2) & 58.5 & 4.3 & 2.4$\times$10$^4$   \\
IRAS 18196$-$1331 & 4 & 18:22:26.80 & $-$13:30:55.00 & 17.629 & 0.149 & (1,1),(2,2) & 22.5 & 2.2 & 6.3$\times$10$^4$    \\
IRAS 18223$-$1243 & 4 & 18:25:10.53 & $-$12:42:25.80 & 18.654 & $-$0.059 & (1,1),(2,2) & 44.9 & 3.5$^{7}$ & 2.2$\times$10$^4$   \\
IRAS 18236$-$1205 & 4 & 18:26:24.30 & $-$12:03:26.80 & 19.369 & $-$0.021 & (1,1),(2,2) & 26.2 & 2.4$^{7}$ & 6.3$\times$10$^3$   \\
IRAS 18264$-$1152 & 2 & 18:29:14.32 & $-$11:50:25.55 & 19.883 & $-$0.534 & (1,1),(2,2) & 44.2 & 3.3 & 1.3$\times$10$^4$   \\
IRAS 18272$-$1217 & 4 & 18:30:02.86 & $-$12:15:16.90 & 19.608 & $-$0.901 & (1,1),(2,2) & 34.0 & 2.8$^{5}$ & 1.5$\times$10$^4$   \\
IRAS 18290$-$0924 & 5 & 18:31:43.94 & $-$09:22:13.30 & 22.356 & 0.067 & (1,1),(2,2) & 84.4 & 4.9$^{8}$ & 2.8$\times$10$^4$    \\
IRAS 18308$-$0841 & 5 & 18:33:33.05 & $-$08:39:10.10 & 23.200 & $-$0.000 & (1,1),(2,2) & 77.1 & 4.6$^{7}$ & 2.0$\times$10$^4$    \\ 
IRAS 18310$-$0825 & 5 & 18:33:47.90 & $-$08:23:45.60 & 23.456 & 0.064 & (1,1),(2,2) & 83.8 & 4.8$^{7}$ & 1.6$\times$10$^4$    \\
IRAS 18337$-$0743 & 5 & 18:36:40.90 & $-$07:39:39.90 & 24.437 & $-$0.231 & (1,1),(2,2) & 60.4 & 3.8$^{7}$ & 2.1$\times$10$^4$    \\
IRAS 18345$-$0641 & 3 & 18:37:17.00 & $-$06:38:32.49 & 25.410 & 0.104 & (1,1),(2,2) & 95.3 & 5.2 & 1.5$\times$10$^4$    \\
IRAS 18352$-$0148 & 5 & 18:37:50.50 & $-$01:45:47.90 & 29.811 & 2.219 & (1,1),(2,2) & 44.8 & 3.0$^{5}$ & 5.1$\times$10$^3$    \\
IRAS 18360$-$0537 & 5 & 18:38:40.30 & $-$05:35:06.40 & 26.508 & 0.283 & (1,1),(2,2) & 101.7 & 5.5$^{7}$ & 6.5$\times$10$^4$    \\
IRAS 18361$-$0627 & 5 & 18:38:57.00 & $-$06:25:00.70 & 25.801 & $-$0.160 & (1,1),(2,2) & 95.5 & 5.2$^{7}$ & 4.8$\times$10$^4$    \\
IRAS 18372$-$0541 & 5 & 18:39:56.00 & $-$05:38:47.00 & 26.597 & $-$0.024 & (1,1),(2,2) & 22.4 & 13.2$^{8,9}$ & 2.6$\times$10$^5$    \\
IRAS 18414$-$0339 & 5 & 18:44:06.90 & $-$03:36:45.90 & 28.882 & $-$0.020 & (1,1),(2,2) & 101.0 & 5.6 & 1.3$\times$10$^4$    \\
IRAS 18426$-$0204 & 4 & 18:45:11.85 & $-$02:01:14.10 & 30.422 & 0.466 & (1,1),(2,2) & 15.0 & 1.2 & 1.1$\times$10$^3$ \\
IRAS 18447$-$0229 & 4 & 18:47:21.60 & $-$02:26:06.20 & 30.300 & $-$0.204 & (1,1),(2,2) & 102.6 & 5.8 & 3.4$\times$10$^4$    \\
IRAS 18460$-$0307 & 3 & 18:48:38.43 & $-$03:03:53.51 & 29.885 & $-$0.775 & (1,1),(2,2) & 84.9 & 4.8$^{7}$ & 1.6$\times$10$^4$    \\
IRAS 18470$-$0044 & 4 & 18:49:37.80 & $-$00:41:01.10 & 32.117 & 0.091 & (1,1) & 89.1 & 9.1$^{8}$ & 1.2$\times$10$^5$ \\
IRAS 18472$-$0022 & 4 & 18:49:52.40 & $-$00:18:56.00 & 32.473 & 0.204 & (1,1),(2,2) & 50.9 & 10.9$^{9}$ & 1.0$\times$10$^5$    \\
IRAS 18488+0000 & 4 & 18:51:25.60 & +00:04:06.50 & 32.992 & 0.034 & (1,1),(2,2) & 82.7 & 9.2$^{8}$ & 1.1$\times$10$^5$    \\
IRAS 18507+0121 & 6 & 18:53:17.40 & +01:24:55.00 & 34.403 & 0.233 & (1,1),(2,2) & 57.1 & 1.6$^{10}$ & 5.6$\times$10$^3$    \\
IRAS 18511+0146 & 4 & 18:53:42.20 & +01:42:52.80 & 34.716 & 0.278 & \nodata &\nodata &\nodata &\nodata \\
IRAS 18517+0437 & 1 & 18:54:13.79 & +04:41:31.76 & 37.427 & 1.518 & (1,1),(2,2) & 43.3 & 2.8$^{5}$ & 1.6$\times$10$^4$    \\
IRAS 18521+0134 & 4 & 18:54:40.76 & +01:38:05.00 & 34.756 & 0.024 & (1,1),(2,2) & 78.5 & 9.1$^{8}$  & 6.3$\times$10$^4$    \\
IRAS 18530+0215 & 4 & 18:55:33.96 & +02:19:07.30 & 35.466 & 0.139 & (1,1),(2,2) & 75.8 & 4.6$^{7}$ & 7.9$\times$10$^4$    \\
IRAS 18532+0047 & 4 & 18:55:50.70 & +00:51:22.40 & 34.196 & $-$0.590 & (1,1),(2,2) & 59.2 & 3.7$^{5}$ & 8.5$\times$10$^3$    \\
IRAS 18553+0414 & 4 & 18:57:53.40 & +04:18:15.70 & 37.498 & 0.530 & (1,1),(2,2) & 11.2       & 12.3$^{9}$ & 8.6$\times$10$^4$   \\
IRAS 19001+0402 & 4 & 19:02:36.20 & +04:06:58.10 & 37.868 & $-$0.602 & (1,1),(2,2) & 50.1 & 3.2$^{5}$ & 8.4$\times$10$^3$    \\
IRAS 19012+0536 & 4 & 19:03:45.40 & +05:40:42.70 & 39.388 & $-$0.141 & (1,1),(2,2) & 65.2 & 4.2 & 1.9$\times$10$^4$    \\
IRAS 19035+0641 & 4 & 19:06:01.59 & +06:46:42.90 & 40.624 & $-$0.137 & (1,1),(2,2) & 32.4 & 2.3$^{1}$ & 8.2$\times$10$^3$    \\
IRAS 19043+0726 & 4 & 19:06:47.70 & +07:31:37.90 & 41.377 & 0.038 & (1,1),(2,2) & 58.9 & 3.9 & 9.8$\times$10$^3$ \\
IRAS 19074+0752 & 4 & 19:09:53.70 & +07:57:15.60 & 42.110 & $-$0.447 & (1,1),(2,2) & 56.0 & 8.7$^{8}$ & 8.7$\times$10$^4$    \\
IRAS 19092+0841 & 4 & 19:11:37.40 & +08:46:30.20 & 43.035 & $-$0.447 & (1,1) (2,2) & 59.2 & 4.1$^{5}$ & 9.8$\times$10$^3$    \\
IRAS 19217+1651 & 4 & 19:23:58.77 & +16:57:44.80 & 51.679 & 0.720 & (1,1),(2,2) & 4.7 & 9.8$^{1}$ & 9.2$\times$10$^4$    \\
IRAS 19220+1432 & 4 & 19:24:20.05 & +14:38:03.60 & 49.669 & $-$0.457 & (1,1),(2,2) & 70.0 & 5.4$^{3}$ & 4.3$\times$10$^4$    \\
IRAS 19266+1745 & 4 & 19:28:55.72 & +17:52:01.90 & 53.037 & 0.112 & (1,1),(2,2) & 4.4 & 9.5$^{9}$ & 5.0$\times$10$^4$    \\
IRAS 19368+2239 & 4 & 19:38:58.10 & +22:46:32.10 & 58.470 & 0.433 & (1,1),(2,2) & 37.0 & 4.4$^{3}$ & 6.6$\times$10$^3$    \\
IRAS 19474+2637 & 4 & 19:49:32.60 & +26:45:13.90 & 63.116 & 0.340 & (1,1),(2,2) & 20.4 & 3.1 & 3.6$\times$10$^3$    \\
IRAS 20050+2720 & 4 & 20:07:06.80 & +27:28:52.80 & 65.780 & $-$2.612 & (1,1),(2,2) & 4.6 & 0.7$^{4}$ & 2.7$\times$10$^2$    \\
IRAS 20099+3640 & 4 & 20:11:46.40 & +36:49:36.60 & 74.161 & 1.644 & \nodata &\nodata &\nodata &\nodata \\
IRAS 20278+3521 & 4 & 20:29:46.90 & +35:31:38.90 & 75.154 & $-$2.085 & (1,1) & $-$3.3 & 4.1$^{11}$ & 6.4$\times$10$^3$ \\
IRAS 20286+4105 & 4 & 20:30:28.00 & +41:15:47.60 & 79.873 & 1.179 & (1,1),(2,2) & $-$5.0 & 3.0$^{3}$ & 2.0$\times$10$^4$    \\
IRAS 20332+4124 & 4 & 20:34:59.73 & +41:34:49.40 & 80.633 & 0.684 & (1,1),(2,2) & $-$2.0 & 1.8$^{11}$ & 8.0$\times$10$^3$    \\
IRAS 23033+5951 & 5 & 23:05:25.31 & +60:08:06.30 & 110.093 & $-$0.067 & (1,1),(2,2) & $-$53.7 & 4.4$^{3}$ & 3.2$\times$10$^4$    \\
IRAS 23139+5939 & 5 & 23:16:10.46 & +59:55:28.50 & 111.256 & $-$0.770 & (1,1),(2,2) & $-$44.1 & 3.5$^{3}$ & 1.3$\times$10$^4$    \\
IRAS 23545+6508 & 5 & 23:57:05.70 & +65:24:50.40 & 117.315 & 3.137 & (1,1) & $-$17.8 & 1.1$^{3}$ & 2.6$\times$10$^3$
\enddata
\tablenotetext{a}{Project indices are defined in Table \ref{obssummary}.}
\tablenotetext{b}{Coordinates are phase centers of the VLA observations.}
\tablenotetext{c}{Distance references: 1. \cite{sridharan2002} 2. \cite{niinuma2011} 3. At the tangential point or in the second or third quadrant of the Milky Way. 4. \cite{wilking1989} 5. \cite{fish2003} 6. \cite{sato2010} 7. IRDC candidates. 8 \cite{anderson2009} 9. \cite{fazal2008} 10. \cite{kurayama2011} 11. The far distance is applied because the derived near distance is too small ($\leq$1 kpc).} 
\end{deluxetable}

\begin{deluxetable}{ccccccrrccc}
\tabletypesize{\scriptsize}
\tablecaption{Derived physical properties.\label{physical}}
\tablewidth{0pt}
\tablehead{
\multirow{2}{*}{Source Name} & {\it $\Delta$v} & $\tau(1,1,m)$\tablenotemark{a} & {\it T$_R$} & {\it T$_K$} &
{\it N}($\rm {NH_3}$) & Mass & Dust Mass\tablenotemark{b} & \multirow{2}{*}{Masers\tablenotemark{c}} & \multirow{2}{*}{Maser Refs\tablenotemark{d}} & \multirow{2}{*}{Morph.\tablenotemark{e}} \\
 & (km/s) & & (K) & (K) & (10$^{13}$cm$^{-2}$) & (M$_{\odot}$) & (M$_{\odot}$) & & &
}
\startdata
IRAS 05358+3543 & 0.01-3.8 & 0.001-2.8 & 8-38 & 8-57 & 0.3-530 & 48 & 386 & m, h ,w & 1, 2, 3 & \nodata \\
IRAS 05490+2658 & 0.01-4.7 & 0.001 & 8-25 & 8-30 & 0.2-220 & 4 & 180 & \nodata & \nodata & f\\
IRAS 06061+2151 & 0.01-2.2 & 0.001 & 8-35 & 8-52 & 0.3-63 & 4 & \nodata & m, w & 4, 5 & \nodata\\
IRAS 06103+1523 & 0.05-2.2 & 0.001 & 8-33 & 8-46 & 2-130 & 25 & \nodata & w & 6 & \nodata\\
IRAS 06382+0939 & 0.01-2.7 & 0.001-4.8 & \nodata & \nodata & 0.2-210 & 2 & \nodata & \nodata & \nodata & d\\
IRAS 18048$-$2019 & 0.05-7.0 & 0.001-4.9 & 8-33 & 8-47 & 2-3100 & 935 & \nodata & m, w & 7, 8 & \nodata\\
IRAS 18089$-$1732 & 0.16-5.1 & 0.001-4.7 & 8-29 & 8-38 & 8-4300 & 1897 & 1200 & m, h, w & 9, 2, 8 & c\\
IRAS 18090$-$1832 & 0.05-3.2 & 0.001-2.4 & 8-24 & 8-29 & 2-520 & 235 & 1223 & m, h & 3, 10 &\nodata \\
W33A & 0.01-7.2 & 0.001-10 & 8-67 & 8-230 & 0.4-9900 & 1938 & \nodata & m, h, w & 9, 2, 11 & f\\
IRAS 18151$-$1208 & 0.01-2.0 & 0.001-2.5 & 8-30 & 8-40 & 0.2-760 & 136 & 481 &m, w & 9, 3 &\nodata \\
IRAS 18159$-$1550 & 0.01-4.2 & 0.001-3.4 & 8-28 & 8-36 & 0.3-650 & 364 & 360 &\nodata & \nodata &\nodata \\
IRAS 18159$-$1648 & 0.01-4.0 & 0.001-10 & 8-40 & 8-63 & 0.3-1700 & 148 & \nodata &m, w & 12, 13 & f\\
IRAS 18182$-$1433 & 0.01-3.5 & 0.001-3.4 & 8-32 & 8-44 & 0.3-3200 & 981 & 1376 &m, h, w & 3, 11 & c \\
IRAS 18196$-$1331 & 0.04-4.7 & 0.001-8.9 & 8-34 & 8-49 & 2-590 & 83 &\nodata &m, h, w & 4, 2, 14 & d\\
IRAS 18223$-$1243 & 0.01-4.1 & 0.001-10 & 8-26 & 8-32 & 0.3-1300 & 443 & 472 &\nodata & \nodata & f \\
IRAS 18236$-$1205 & 0.01-3.8 & 0.001-10 & 8-32 & 8-43 & 0.3-2000 & 178 & \nodata &m, w & 7, 8 &\nodata \\
IRAS 18264$-$1152 & 0.26-4.0 & 0.001-3.2 & 8-31 & 8-41 & 26-1700 & 309  &1928 &m, w & 3 &\nodata \\
IRAS 18272$-$1217 & 0.09-3.8 & 0.001-10 & 8-21 & 8-25 & 4-2200 & 70 & 107 &\nodata  & \nodata & \nodata\\
IRAS 18290$-$0924 & 0.01-2.1 & 0.001-4.6 & 8-24 & 8-30 & 0.3-910 & 1182 & 1660 &m, w & 3  &\nodata\\
IRAS 18308$-$0841 & 0.01-2.9 & 0.001-10 & 8-33 & 8-45 & 0.3-1800 & 2186 & 1773 &w & 3 & f\\
IRAS 18310$-$0825 & 0.04-2.1 & 0.001-6.0 & 8-37 & 8-57 & 1-970 & 1312 & 1631 &m & 3 & f\\
IRAS 18337$-$0743 & 0.01-4.6 & 0.001-4.4 & 8-28 & 8-36 & 0.2-1600 & 2657 & 1230 &\nodata & \nodata & f\\
IRAS 18345$-$0641 & 0.01-3.9 & 0.001-2.9 & 8-30 & 8-39 & 0.3-1300 & 1234 & 1028 & m & 3 &\nodata \\
IRAS 18352$-$0148 & 0.01-2.5 & 0.001-2.8 & 8-18 & 8-20 & 0.3-500 & 128 & \nodata &\nodata & \nodata & f\\
IRAS 18360$-$0537 & 0.01-4.4 & 0.001-10 & 8-50 & 8-101 & 0.2-6500 & 4746 & \nodata &w & 6, 8 & c\\
IRAS 18361$-$0627 & 0.01-2.4 & 0.001-2.6 & 8-46 & 8-84 & 0.2-440 & 173 & \nodata & m & 4 & \nodata\\
IRAS 18372$-$0541 & 0.01-3.9 & 0.001-10 & 8-42 & 8-71 & 0.8-2400 & 4432 & 4948 &m, w & 3 & f\\
IRAS 18414$-$0339 & 0.07-6.5 & 0.001-10 & 8-25 & 8-31 & 2-1400 & 1043  &\nodata &\nodata & \nodata & c\\
IRAS 18426$-$0204 & 0.05-3.9 & 0.001-2.4 & 8-20 & 8-23 & 1-270 & 4 & 43 &\nodata & \nodata &\nodata \\
IRAS 18447$-$0229 & 0.01-2.5 & 0.001-6.2 & 8-25 & 8-30 & 0.4-480 & 344 & 523 &\nodata & \nodata  &\nodata\\
IRAS 18460$-$0307 & 0.01-2.5 & 0.001-5.8 & 8-29 & 8-37 & 0.2-990 & 533 & 771 &\nodata & \nodata & f\\
IRAS 18470$-$0044 & 0.05-2.4 & 0.001 & \nodata & \nodata & 2-130 & 129 & 2581 & \nodata & \nodata &\nodata \\
IRAS 18472$-$0022 & 0.04-3.9 & 0.001-5.9 & 8-42 & 8-70 & 0-2000 & 1960 & 5775 &\nodata & \nodata & f\\
IRAS 18488+0000 & 0.01-3.9 & 0.001-10 & 8-28 & 8-35 & 0.8-1400 & 1139  & 3853 &m, w & 15, 3 &\nodata \\
IRAS 18507+0121 & 0.1-5.5 & 0.001-7.3 & 8-30 & 8-40 & 3-2800 & 309 & \nodata &m, w & 4, 16 & f\\
IRAS 18517+0437 & 0.21-3.6 & 0.001-1.8 & 8-31 & 8-42 & 10-600 & 91 & 1078 & m, w & 17, 8 & \nodata\\
IRAS 18521+0134 & 0.05-3.2 & 0.001-3.8 & 8-37 & 8-57 & 1-1500 & 1537 & 1532 & m & 3 &\nodata \\
IRAS 18530+0215 & 0.01-3.5 & 0.001-10 & 8-36 & 8-52 & 0.5-1000 & 636 & 1741 &\nodata & \nodata & f\\
IRAS 18532+0047 & 0.01-4.9 & 0.001-3.9 & 8-26 & 8-33 & 0.4-1700 & 286 & \nodata &\nodata & \nodata &\nodata \\
IRAS 18553+0414 & 0.08-4.3 & 0.001-0.2 & 8-27 & 8-35 & 2-220 & 353 & 7194 &w & 3 &\nodata  \\
IRAS 19001+0402 & 0.04-4.5 & 0.001-10 & 8-30 & 8-40 & 0.9-670 & 95 & \nodata &\nodata & \nodata & \nodata\\
IRAS 19012+0536 & 0.05-4.6 & 0.001-2.0 & 8-35 & 8-50 & 2-1300 & 132 & 463 &w & 3 &\nodata \\
IRAS 19035+0641 & 0.05-5.0 & 0.001-2.4 & 8-26 & 8-33 & 1-1600 & 98 & 211 &m, h, w & 3, 2, 11 &\nodata \\
IRAS 19043+0726 & 0.05-3.2 & 0.001-0.7 & 8-32 & 8-45 & 2-340 & 26 & \nodata &\nodata & \nodata &\nodata \\
IRAS 19074+0752 & 0.01-2.3 & 0.001-1.8 & 8-27 & 8-34 & 0.4-230 & 296 & 1032 &\nodata & \nodata & f\\
IRAS 19092+0841 & 0.1-3.8 & 0.001-3.7 & 8-33 & 8-46 & 2-1700 & 241 & \nodata &m, h, w & 18, 19, 8 &\nodata \\
IRAS 19217+1651 & 0.05-3.2 & 0.001-10 & 8-28 & 8-35 & 1-1400 & 1903 & 4146 &m, w & 3 &\nodata \\
IRAS 19220+1432 & 0.05-4.1 & 0.001-5.1 & 8-20 & 8-24 & 2-1100 & 308 & 1642 &\nodata & \nodata & f\\
IRAS 19266+1745 & 0.04-3.5 & 0.001-2.0 & 8-30 & 8-39 & 2-900 & 996 & 2826 &\nodata & \nodata & f\\
IRAS 19368+2239 & 0.01-3.1 & 0.001-10 & 8-26 & 8-32 & 0.3-1200 & 405 & \nodata &w & 6 & f \\
IRAS 19474+2637 & 0.05-2.8 & 0.001-1.7 & 8-23 & 8-28 & 2-590 & 131 & \nodata & w & 6 & \nodata\\
IRAS 20050+2720 & 0.01-3.6 & 0.001-10 & 8-31 & 8-42 & 0.3-640 & 10 & \nodata &w & 6, 8 & f\\
IRAS 20278+3521 & 0.01-1.9 & 0.001-1.8 & \nodata & \nodata & 0.4-200 & 47 & \nodata & w & 6 & \nodata\\
IRAS 20286+4105 & 0.01-3.5 & 0.001-2.2 & 8-37 & 8-56 & 0.3-640 & 133 & \nodata &w & 8  &\nodata \\
IRAS 20332+4124 & 0.1-5.3 & 0.001-10 & 8-24 & 8-29 & 3-2900 & 74 & 163 &w & 6 &\nodata \\
IRAS 23033+5951 & 0.1-2.8 & 0.001-2.2 & 8-29 & 8-37 & 4-420 & 227 & 1839 &w & 3 & f\\
IRAS 23139+5939 & 0.05-3.6 & 0.001-0.7 & 8-31 & 8-41 & 2-200 & 43 & 468 &m, w & 4, 3  &\nodata \\
IRAS 23545+6508 & 0.01-1.1 & 0.001 & \nodata & \nodata & 0.7-86 & 0.3 & 10 &\nodata & \nodata & \nodata
\enddata
\tablenotetext{a}{0.001 and 10 are the lower and upper limits of $\tau(1,1,m)$.}
\tablenotetext{b}{The masses are taken from Table 3 of \cite{beuther2002cs}, corrected for the emissivity problem noted in the erratum of the paper, and rescaled with the distances we currently use.}
\tablenotetext{c}{m, h, w stand for methanol (both Class I and Class II), hydroxyl and water maser detections, respectively.}
\tablenotetext{d}{References: 1. \citet{minier2000} 2. \citet{argon2000} 3. \citet{beuther2002maser} 4. \citet{szymczak2000} 5. \citet{niinuma2011} 6. \citet{harju1998} 7. \citet{walsh1998} 8. \citet{palla1991} 9. \citet{pestalozzi2005} 10. \citet{caswell1998} 11. \citet{forster1989} 12. \citet{valtts2000} 13. \citet{sato2010} 14. \citet{menten2004} 15. \citet{bartkiewicz2009} 16. \citet{kurayama2011} 17. \citet{schutte1993} 18. \citet{fontani2010} 19. \citet{edris2011}}
\tablenotetext{e}{f, c, d stand for candidates of filaments, NH$_3$ concentration and NH$_3$ dispersed sources, respectively.}
\end{deluxetable}

\begin{deluxetable}{ccccccc}
\tabletypesize{\scriptsize}
\tablecaption{Physical properties of the dense cores.\label{stat}}
\tablewidth{0pt}
\tablehead{
\multirow{2}{*}{} & {\it R} & {\it $\Delta$v} & {\it T$_R$} & {\it N}($\rm NH_3$) & {\it M$_{core}$} & \multirow{2}{*}{\it M$_{vir}$/M$_{core}$} \\
 & (pc) & (km/s) & (K) &  (10$^{15}$cm$^{-2}$) & (M$_{\odot}$) & 
}
\startdata
{\it mean} & 0.08 & 1.08 & 18.1 &  2.3 & 67 & 1.36 \\
{\it median} & 0.07 & 0.93 & 17.5 & 1.3 & 24 & 0.64
\enddata
\end{deluxetable}

\begin{deluxetable}{ccccccc}
\tabletypesize{\scriptsize}
\tablecaption{Cylinder collapse of filamentary gas.\label{cylinder}}
\tablewidth{0pt}
\tablehead{
\multirow{3}{*}{Source Name} & 
\multirow{2}{*}{$\lambda_{\rm obs}$} & 
\multirow{2}{*}{$(M/l)_{\rm obs}$} & 
\multicolumn{2}{c}{Thermal Support} & 
\multicolumn{2}{c}{Turbulent Support}  \\
 \cline{4-5} \cline{6-7}& & & $\lambda_{\rm max}$ & $(M/l)_{\rm max}$ & 
$\lambda_{\rm max}$ & $(M/l)_{\rm max}$  \\ 
 & (pc) & (M$_{\odot}$ pc$^{-1}$) &(pc) & (M$_{\odot}$ pc$^{-1}$) &(pc) & (M$_{\odot}$ pc$^{-1}$)
 }
\startdata
IRAS 18223$-$1243 & 0.26 & 127 & 0.13 - 0.23 & 13 - 45 & 0.20 - 0.60 & 32 - 290 \\
IRAS 18507+0121    & 0.15 & 121 & 0.13 - 0.25 & 13 - 51 & 0.28 - 0.67 & 64 - 360 \\
IRAS 19074+0752    & 0.30 & 110 & 0.15 - 0.27 & 13 - 42 & 0.32 - 0.63 & 57 - 230 \\
IRAS 19220+1432    & 0.23 & 122 & 0.14 - 0.23 & 13 - 34 & 0.26 - 0.81 & 45 - 420 \\
IRAS 19368+2239    & 0.22 & 172 & 0.13 - 0.24 & 13 - 45 & 0.28 - 0.70 & 60 - 370 \\
IRAS 23033+5951    & 0.19 & 142 & 0.14 - 0.24 & 13 - 39 & 0.35 - 0.59 & 86 - 240
\enddata
\tablecomments{The temperature and linewidth used for theoretical derivation are selected from regions without any sign of star formation, therefore are close to the initial conditions.}
\end{deluxetable}

\clearpage

\appendix

\begin{figure}[!t]
\begin{tabular}{p{9.0cm}p{8.1cm}}
\includegraphics[width=9.0cm]{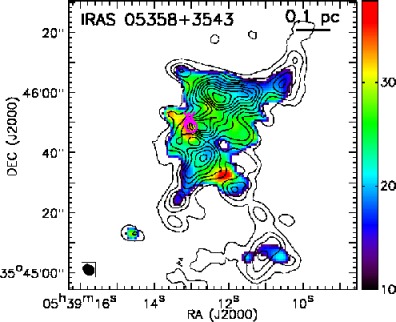} & \includegraphics[width=8.1cm]{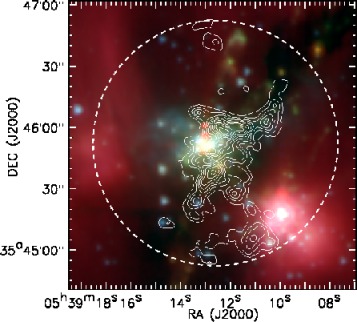} \\
\includegraphics[width=8.1cm]{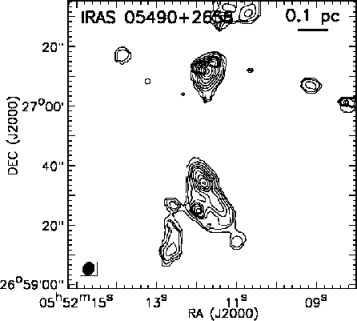} & \includegraphics[width=8.1cm]{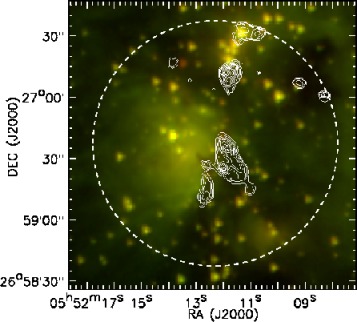} \\
\includegraphics[width=9.0cm]{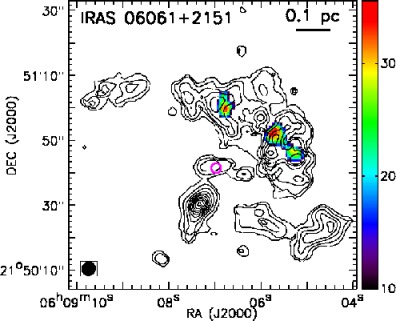} & \includegraphics[width=8.1cm]{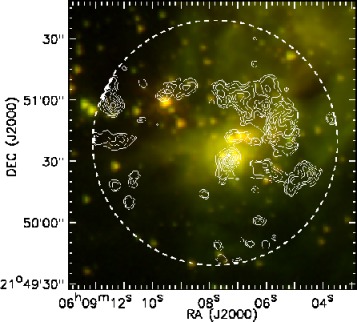} \\
\end{tabular}
\caption{Rotational temperature and IRAC three-color images of all sources in the sample, except IRAS 18182$-$1433, IRAS 18196$-$1331, and IRAS 18223$-$1243 which are presented in Figure \ref{tempfigs}. {\it Left}: Rotational temperatures are presented in color scales. The black contours are the integrated intensities of NH$_3$ (1,1) main hyperfine line, from 5\% to 95\% of the peak intensity in steps of 10\%. Maser detections with interferometers are marked with either circles (water maser), crosses (Class II methanol maser), or triangles (hydroxyl maser). The synthesized beam of the integrated intensities is shown in the lower-left corner in each panel, while the beam of the rotational temperatures should be slightly larger given that the temperatures are derived from the smoothed data. {\it Right}: The Spitzer IRAC composite image, with 8 $\mu$m (red), 4.5 $\mu$m (green) and 3.6 $\mu$m (blue) emission. The NH$_3$ (1,1) integrated intensity contours are identical to those in the left panels. The dashed circle shows the 2$'$ FWHM primary beam of the VLA.}\label{tempfigs_more}
\end{figure}

\clearpage

\addtocounter{figure}{-1}
\begin{figure}[!t]
\begin{tabular}{p{9.0cm}p{8.1cm}}
\includegraphics[width=9.0cm]{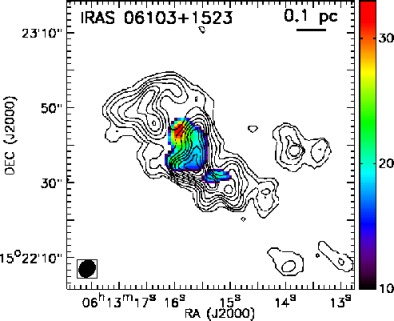} & \includegraphics[width=8.1cm]{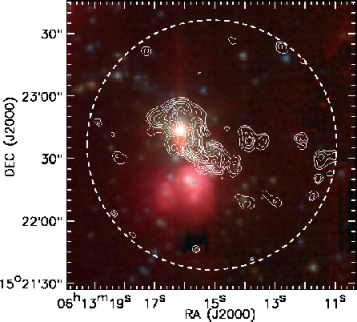} \\
\includegraphics[width=8.15cm]{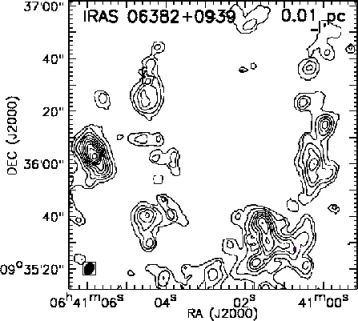} & \includegraphics[width=8.1cm]{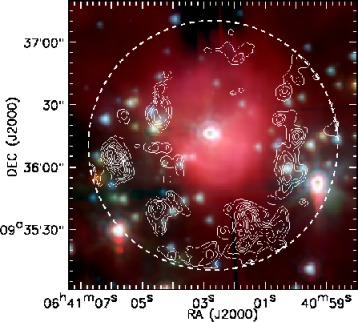} \\
\includegraphics[width=9.0cm]{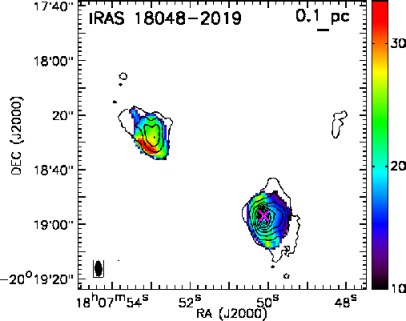} & \includegraphics[width=8.1cm]{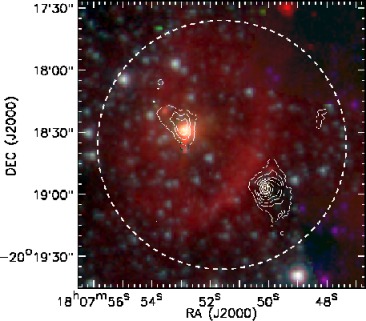} \\
\end{tabular}
\caption{\it Continued}
\end{figure}

\clearpage

\addtocounter{figure}{-1}
\begin{figure}[!t]
\begin{tabular}{p{9.0cm}p{8.1cm}}
\includegraphics[width=9.0cm]{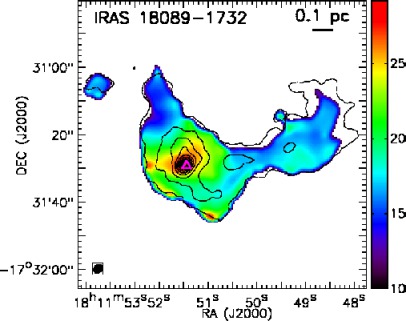} & \includegraphics[width=8.1cm]{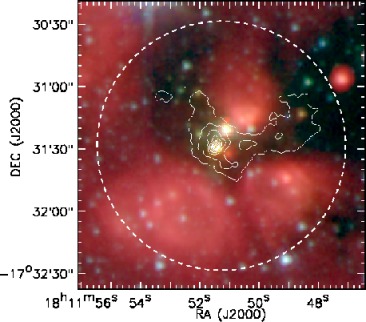} \\
\includegraphics[width=9.0cm]{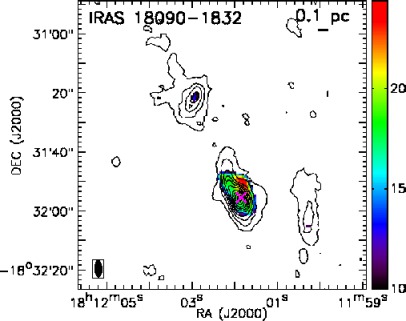} & \includegraphics[width=8.1cm]{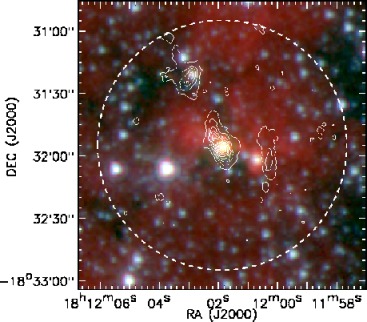} \\
\includegraphics[width=9.0cm]{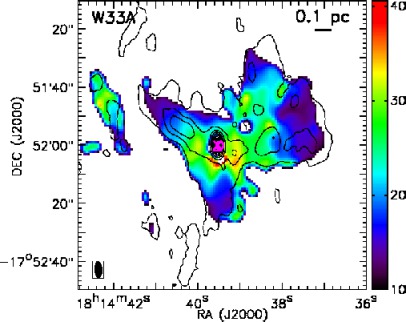} & \includegraphics[width=8.1cm]{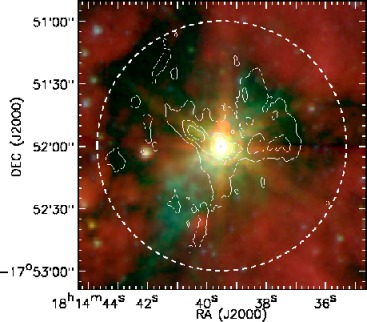} \\
\end{tabular}
\caption{\it Continued}
\end{figure}

\addtocounter{figure}{-1}
\begin{figure}[!t]
\begin{tabular}{p{9.0cm}p{8.1cm}}
\includegraphics[width=9.0cm]{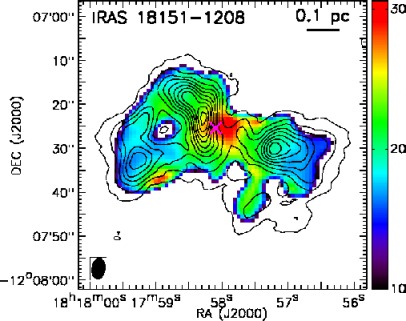} & \includegraphics[width=8.1cm]{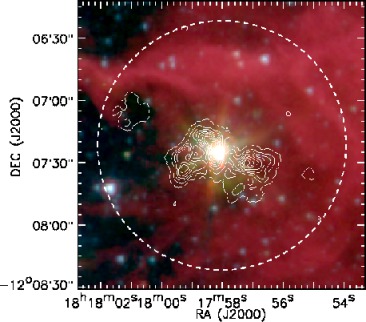} \\
\includegraphics[width=9.0cm]{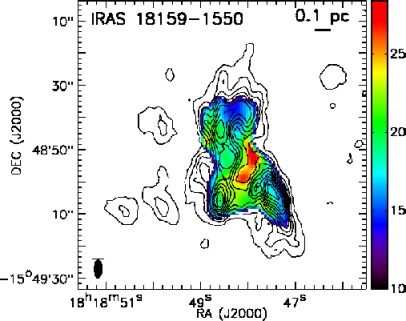} & \includegraphics[width=8.1cm]{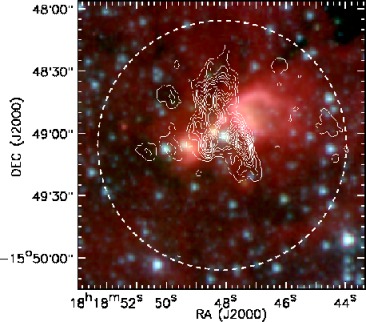} \\
\includegraphics[width=9.0cm]{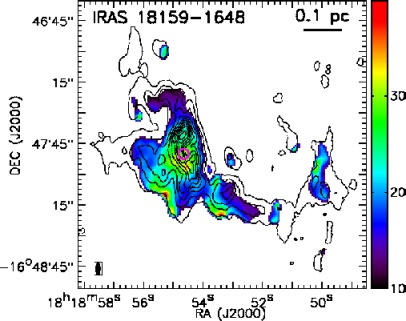} & \includegraphics[width=8.1cm]{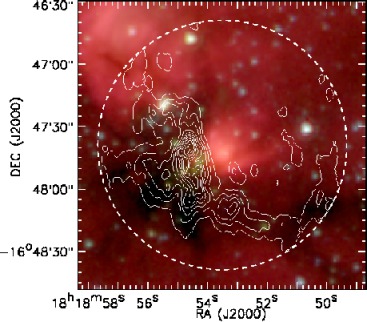} \\
\end{tabular}
\caption{\it Continued}
\end{figure}

\addtocounter{figure}{-1}
\begin{figure}[!t]
\begin{tabular}{p{9.0cm}p{8.1cm}}
\includegraphics[width=9.0cm]{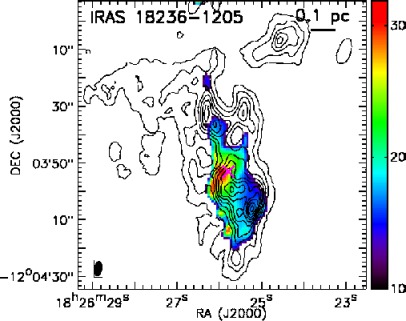} & \includegraphics[width=8.1cm]{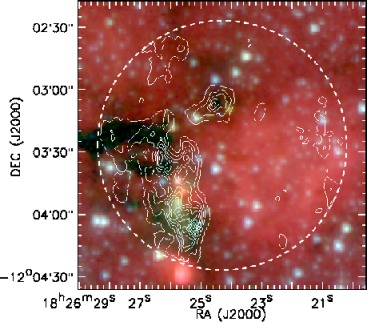} \\
\includegraphics[width=9.0cm]{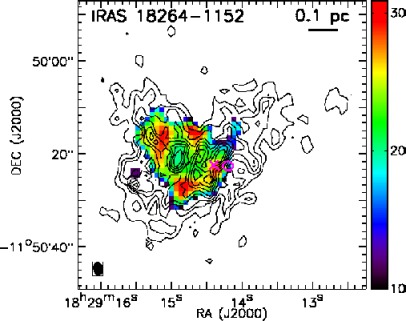} & \includegraphics[width=8.1cm]{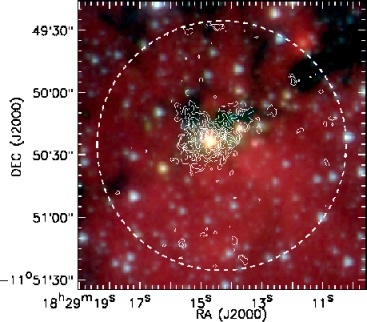} \\
\includegraphics[width=9.0cm]{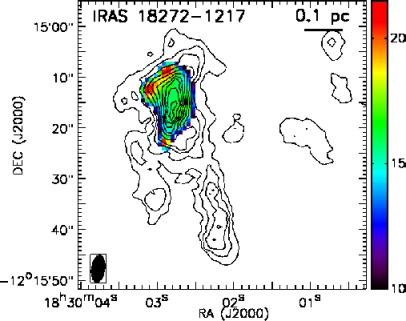} & \includegraphics[width=8.1cm]{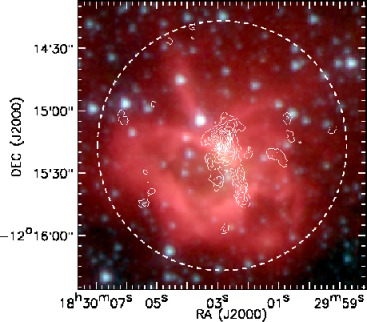} \\
\end{tabular}
\caption{\it Continued}
\end{figure}

\addtocounter{figure}{-1}
\begin{figure}[!t]
\begin{tabular}{p{9.0cm}p{8.1cm}}
\includegraphics[width=9.0cm]{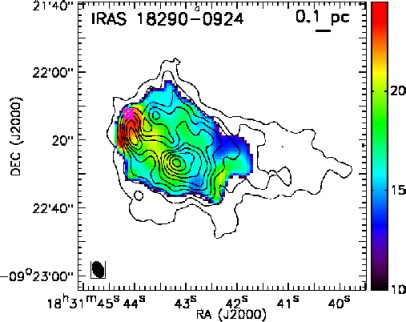} & \includegraphics[width=8.1cm]{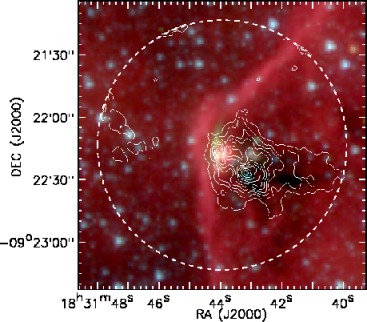} \\
\includegraphics[width=9.0cm]{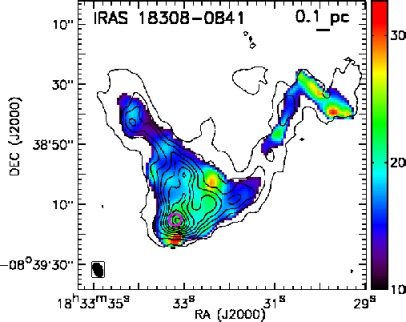} & \includegraphics[width=8.1cm]{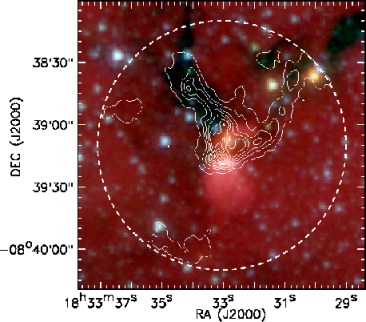} \\
\includegraphics[width=9.0cm]{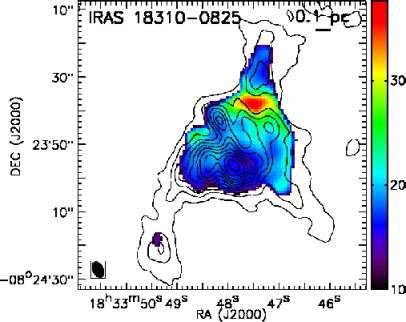} & \includegraphics[width=8.1cm]{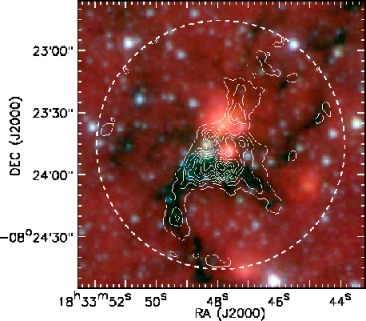} \\
\end{tabular}
\caption{\it Continued}
\end{figure}

\addtocounter{figure}{-1}
\begin{figure}[!t]
\begin{tabular}{p{9.0cm}p{8.1cm}}
\includegraphics[width=9.0cm]{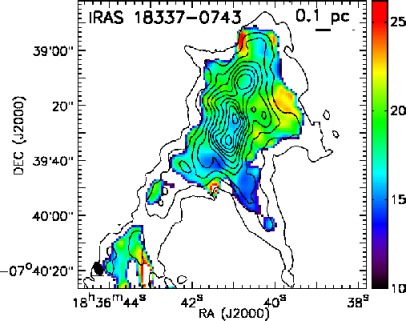} & \includegraphics[width=8.1cm]{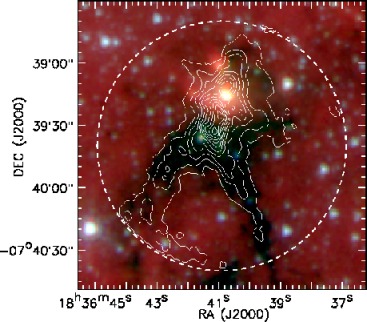} \\
\includegraphics[width=9.0cm]{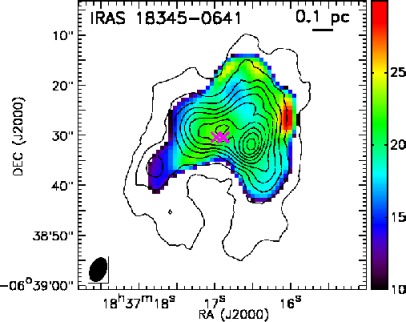} & \includegraphics[width=8.1cm]{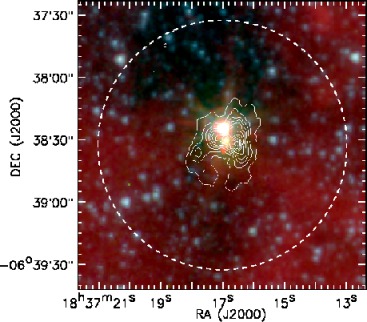} \\
\includegraphics[width=9.0cm]{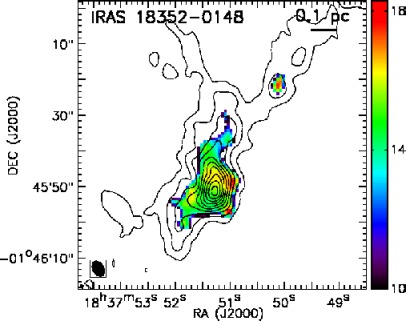} & \includegraphics[width=8.1cm]{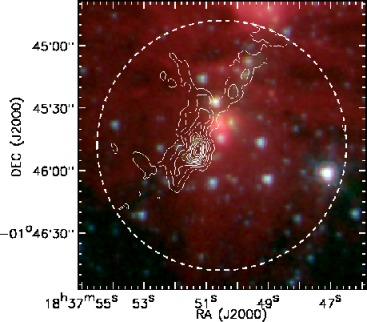} \\
\end{tabular}
\caption{\it Continued}
\end{figure}

\addtocounter{figure}{-1}
\begin{figure}[!t]
\begin{tabular}{p{9.0cm}p{8.1cm}}
\includegraphics[width=9.0cm]{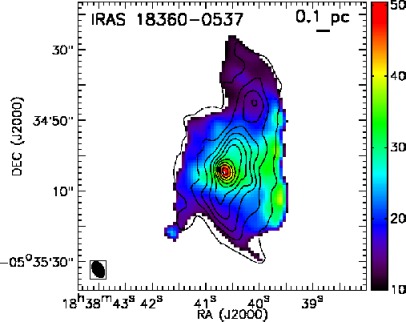} & \includegraphics[width=8.1cm]{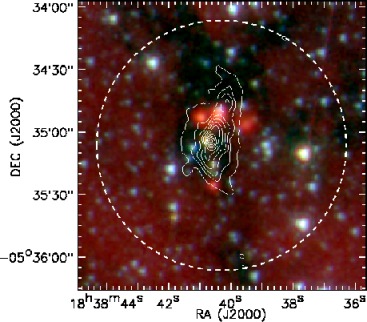} \\
\includegraphics[width=9.0cm]{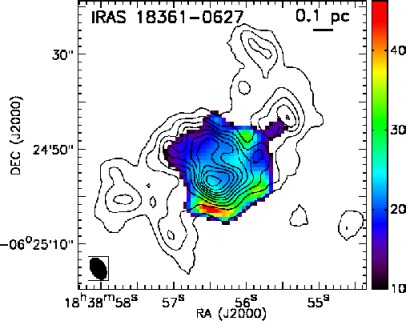} & \includegraphics[width=8.1cm]{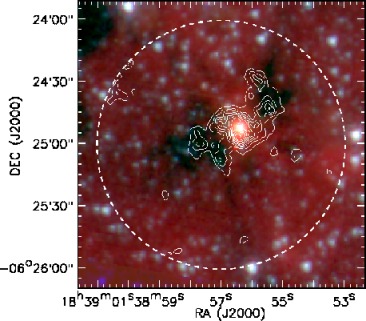} \\
\includegraphics[width=9.0cm]{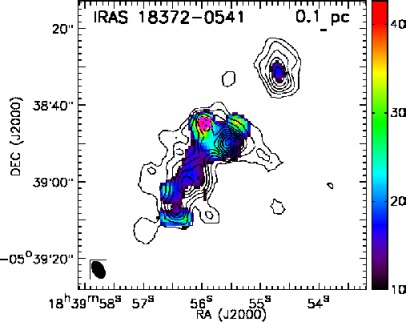} & \includegraphics[width=8.1cm]{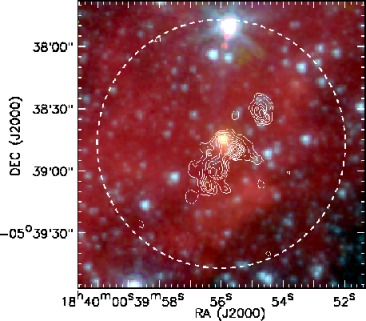} \\
\end{tabular}
\caption{\it Continued}
\end{figure}

\addtocounter{figure}{-1}
\begin{figure}[!t]
\begin{tabular}{p{9.0cm}p{8.1cm}}
\includegraphics[width=9.0cm]{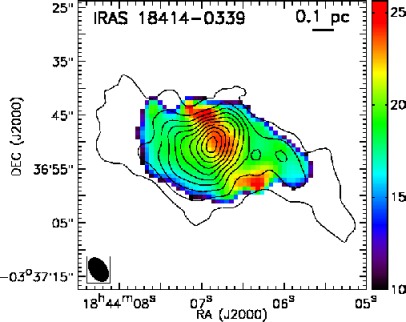} & \includegraphics[width=8.1cm]{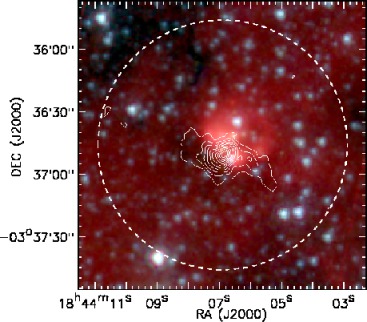} \\
\includegraphics[width=9.0cm]{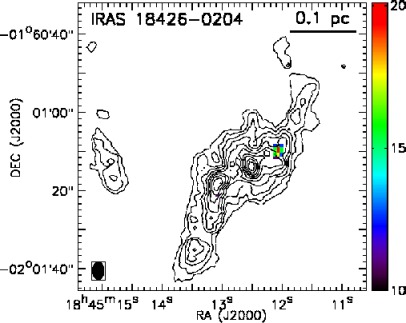} & \includegraphics[width=8.1cm]{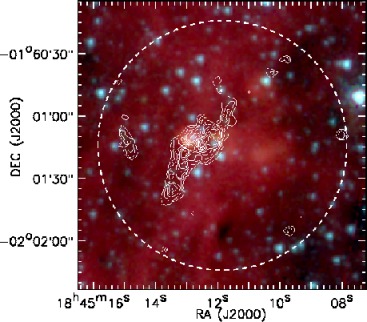} \\
\includegraphics[width=9.0cm]{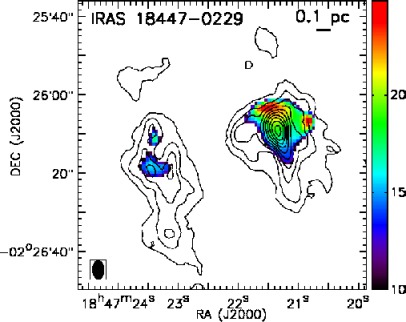} & \includegraphics[width=8.1cm]{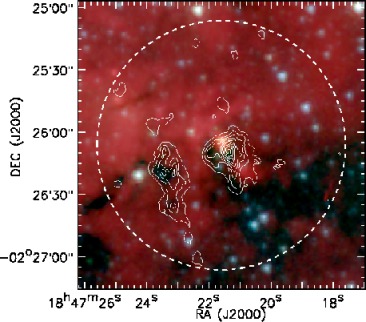} \\
\end{tabular}
\caption{\it Continued}
\end{figure}

\addtocounter{figure}{-1}
\begin{figure}[!t]
\begin{tabular}{p{9.0cm}p{8.1cm}}
\includegraphics[width=9.0cm]{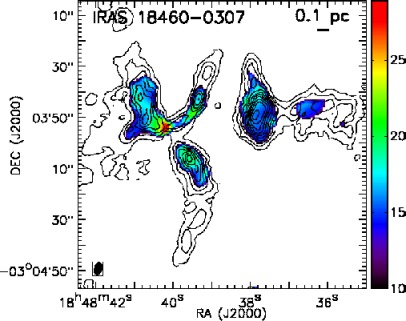} & \includegraphics[width=8.1cm]{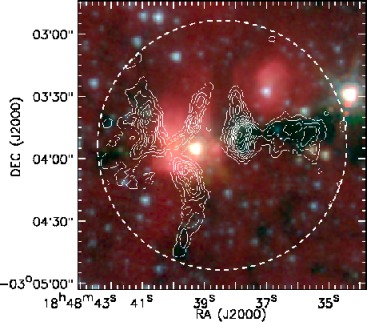} \\
\includegraphics[width=8.15cm]{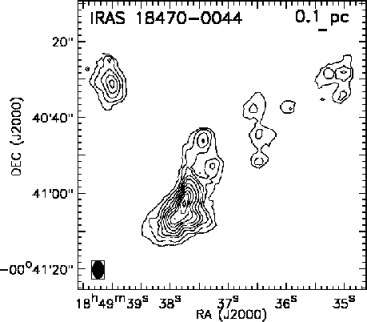} & \includegraphics[width=8.1cm]{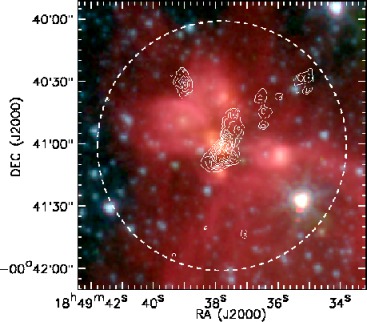} \\
\includegraphics[width=9.0cm]{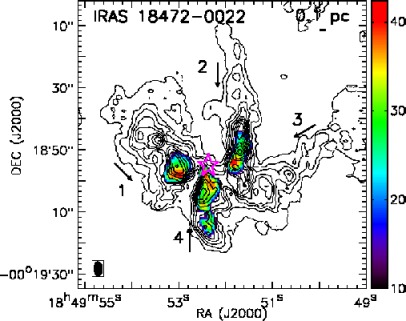} & \includegraphics[width=8.1cm]{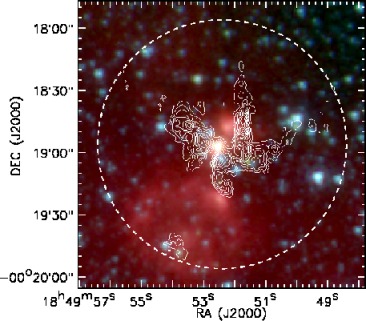} \\
\end{tabular}
\caption{\it Continued}
\end{figure}

\addtocounter{figure}{-1}
\begin{figure}[!t]
\begin{tabular}{p{9.0cm}p{8.1cm}}
\includegraphics[width=9.0cm]{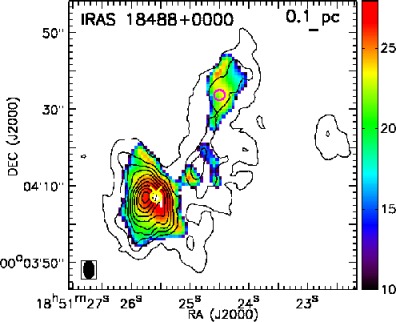} & \includegraphics[width=8.1cm]{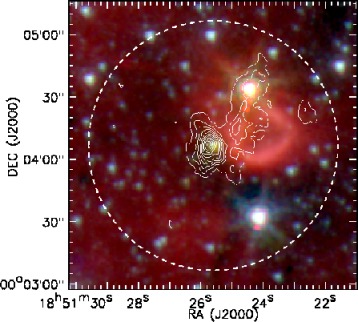} \\
\includegraphics[width=9.0cm]{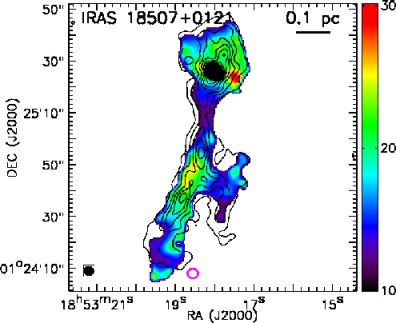} & \includegraphics[width=8.1cm]{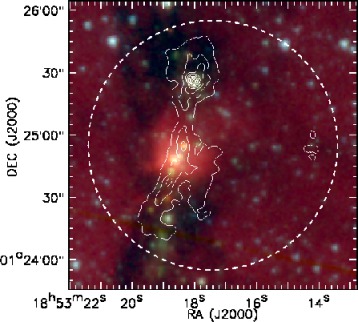} \\
\includegraphics[width=9.0cm]{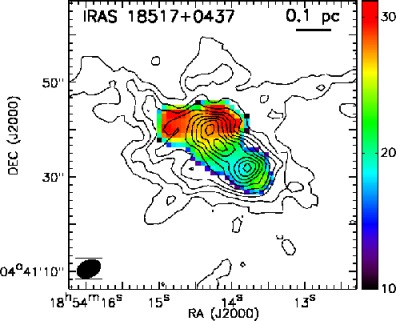} & \includegraphics[width=8.1cm]{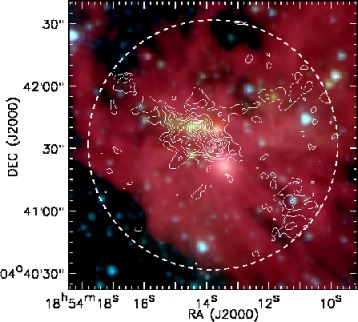} \\
\end{tabular}
\caption{\it Continued}
\end{figure}

\addtocounter{figure}{-1}
\begin{figure}[!t]
\begin{tabular}{p{9.0cm}p{8.1cm}}
\includegraphics[width=9.0cm]{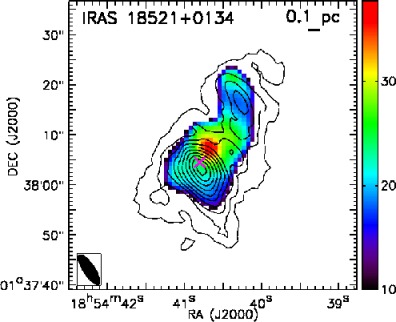} & \includegraphics[width=8.1cm]{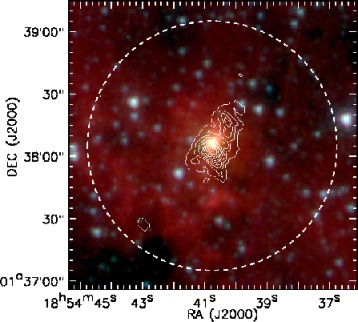} \\
\includegraphics[width=9.0cm]{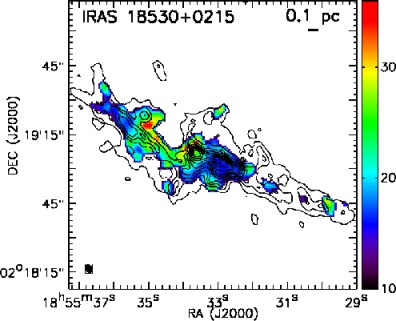} & \includegraphics[width=8.1cm]{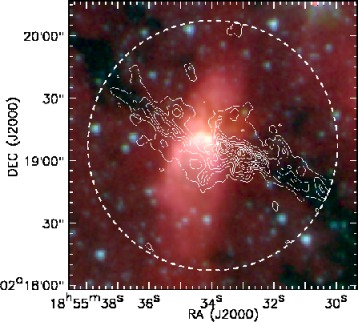} \\
\includegraphics[width=9.0cm]{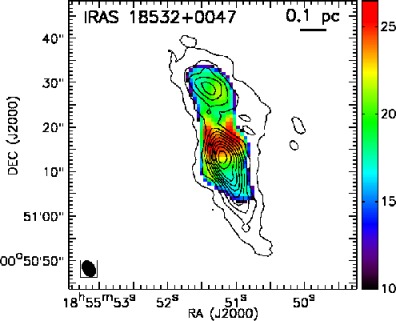} & \includegraphics[width=8.1cm]{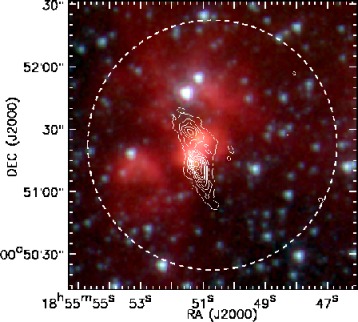} \\
\end{tabular}
\caption{\it Continued}
\end{figure}

\addtocounter{figure}{-1}
\begin{figure}[!t]
\begin{tabular}{p{9.0cm}p{8.1cm}}
\includegraphics[width=9.0cm]{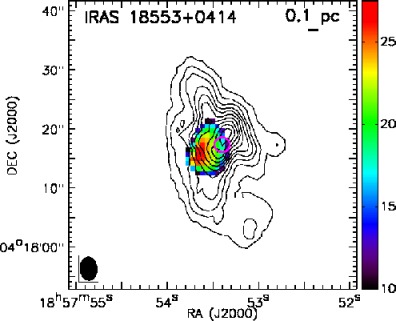} & \includegraphics[width=8.1cm]{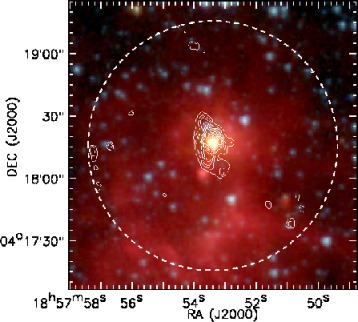} \\
\includegraphics[width=9.0cm]{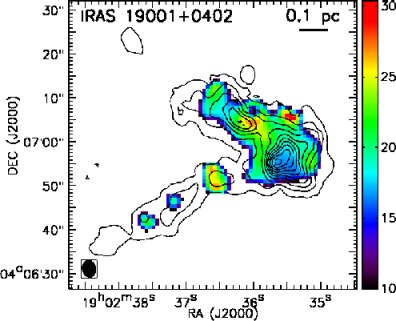} & \includegraphics[width=8.1cm]{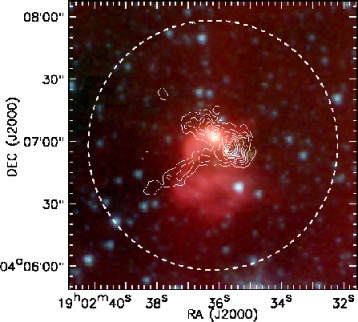} \\
\includegraphics[width=9.0cm]{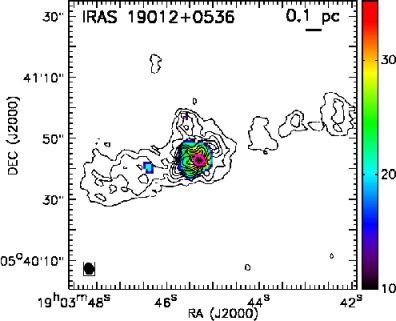} & \includegraphics[width=8.1cm]{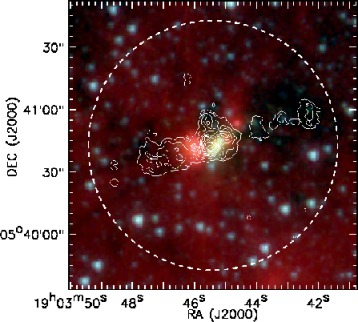} \\
\end{tabular}
\caption{\it Continued}
\end{figure}

\addtocounter{figure}{-1}
\begin{figure}[!t]
\begin{tabular}{p{9.0cm}p{8.1cm}}
\includegraphics[width=9.0cm]{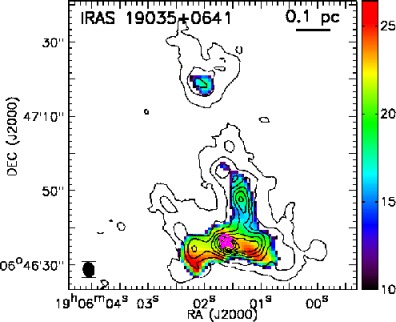} & \includegraphics[width=8.1cm]{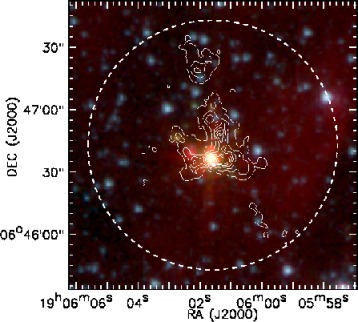} \\
\includegraphics[width=9.0cm]{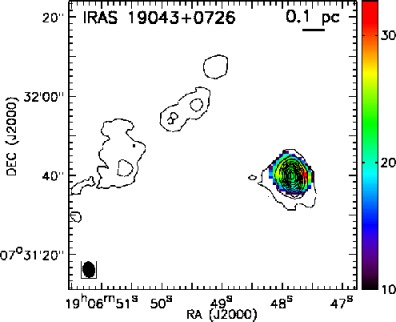} & \includegraphics[width=8.1cm]{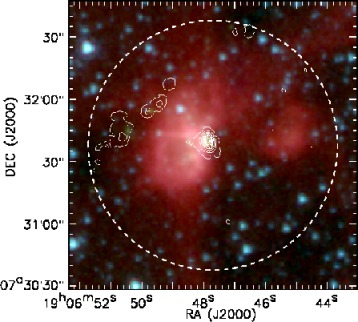} \\
\includegraphics[width=9.0cm]{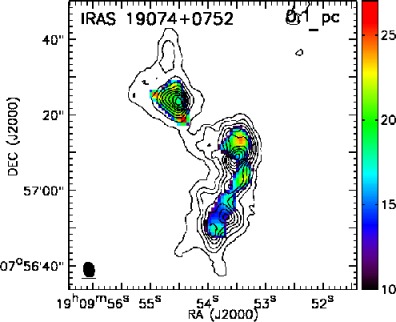} & \includegraphics[width=8.1cm]{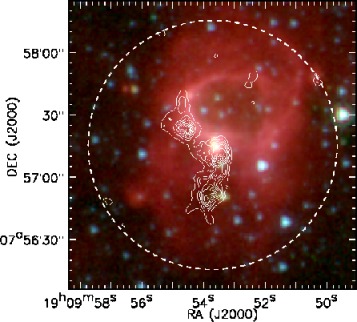} \\
\end{tabular}
\caption{\it Continued}
\end{figure}

\addtocounter{figure}{-1}
\begin{figure}[!t]
\begin{tabular}{p{9.0cm}p{8.1cm}}
\includegraphics[width=9.0cm]{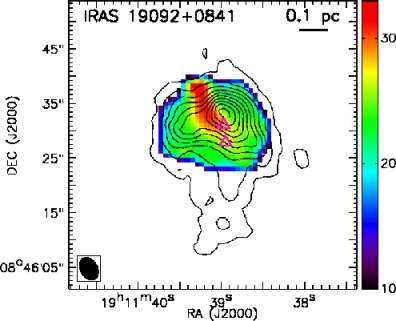} & \includegraphics[width=8.1cm]{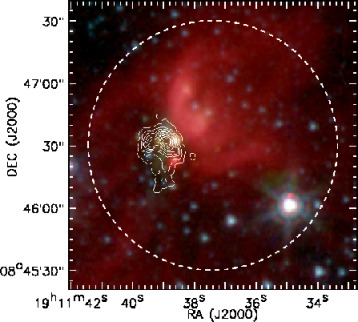} \\
\includegraphics[width=9.0cm]{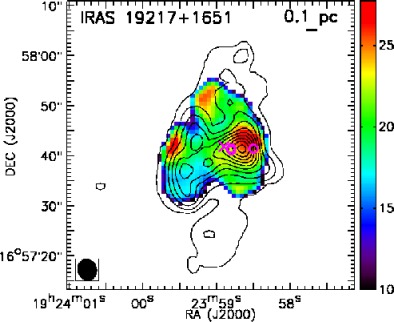} & \includegraphics[width=8.1cm]{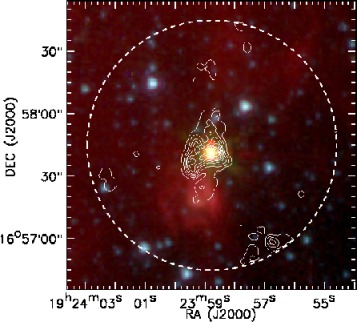} \\
\includegraphics[width=9.0cm]{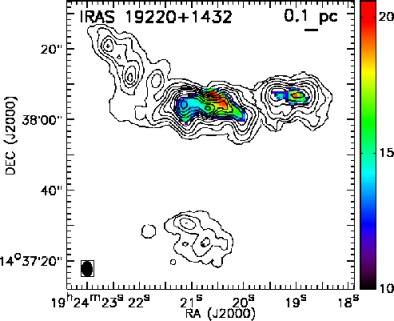} & \includegraphics[width=8.1cm]{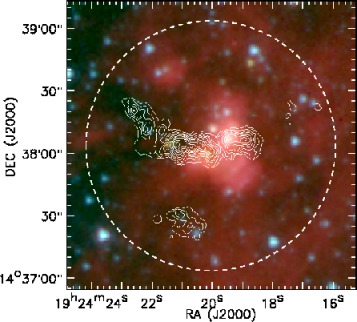} \\
\end{tabular}
\caption{\it Continued}
\end{figure}

\addtocounter{figure}{-1}
\begin{figure}[!t]
\begin{tabular}{p{9.0cm}p{8.1cm}}
\includegraphics[width=9.0cm]{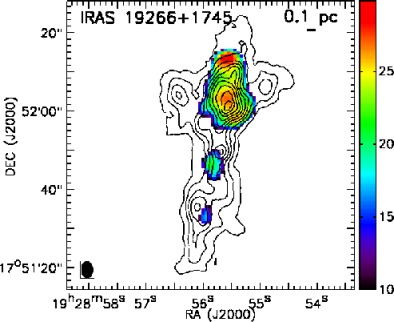} & \includegraphics[width=8.1cm]{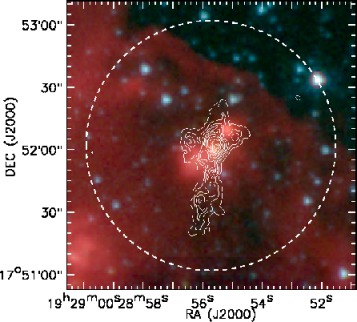} \\
\includegraphics[width=9.0cm]{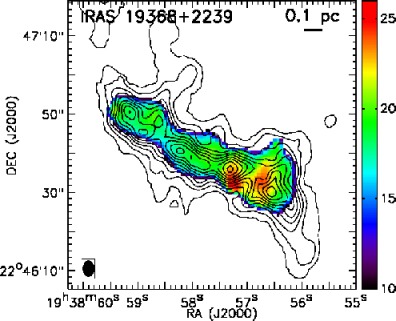} & \includegraphics[width=8.1cm]{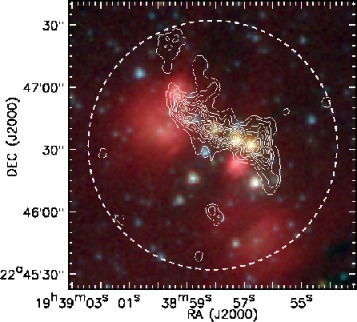} \\
\includegraphics[width=9.0cm]{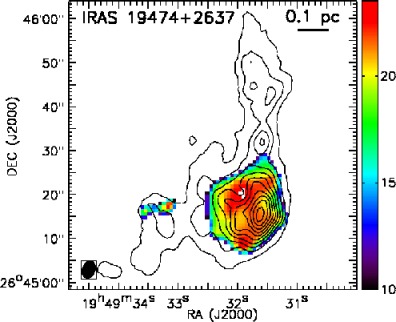} & \includegraphics[width=8.1cm]{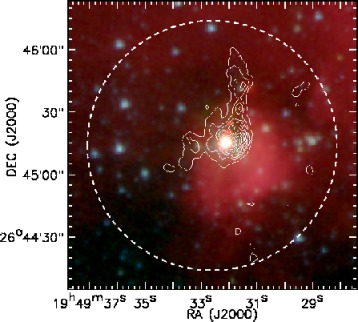} \\
\end{tabular}
\caption{\it Continued}
\end{figure}

\addtocounter{figure}{-1}
\begin{figure}[!t]
\begin{tabular}{p{9.0cm}p{8.1cm}}
\includegraphics[width=9.0cm]{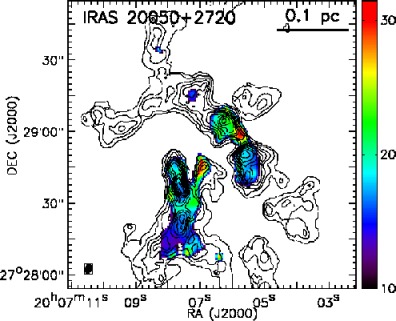} & \includegraphics[width=8.1cm]{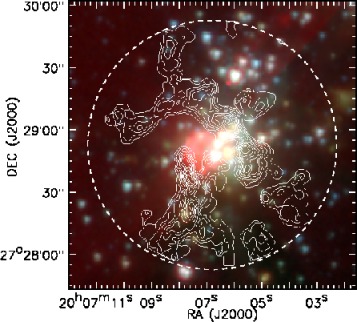} \\
\includegraphics[width=8.15cm]{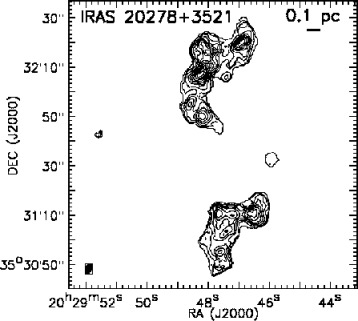} & \includegraphics[width=8.1cm]{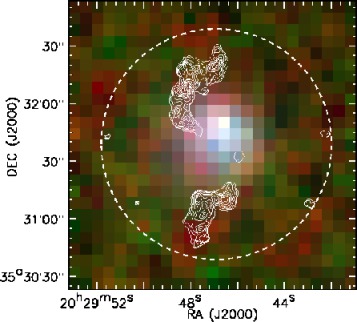} \\
\includegraphics[width=9.0cm]{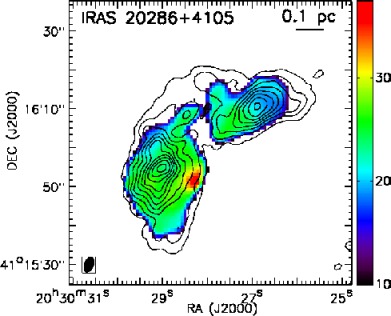} & \includegraphics[width=8.1cm]{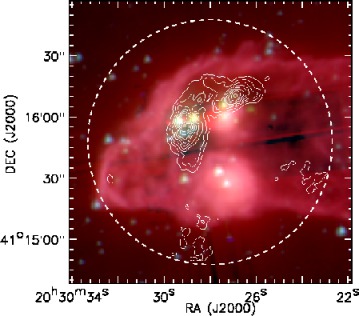} \\
\end{tabular}
\caption{\it Continued}
\end{figure}

\addtocounter{figure}{-1}
\begin{figure}[!t]
\begin{tabular}{p{9.0cm}p{8.1cm}}
\includegraphics[width=9.0cm]{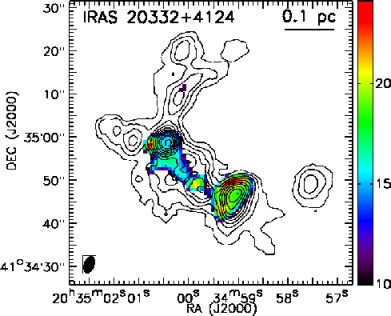} & \includegraphics[width=8.1cm]{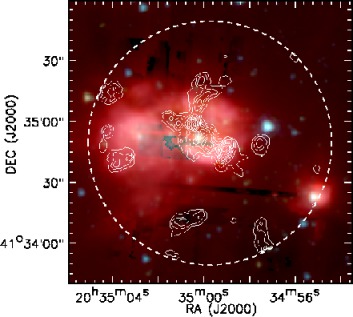} \\
\includegraphics[width=9.0cm]{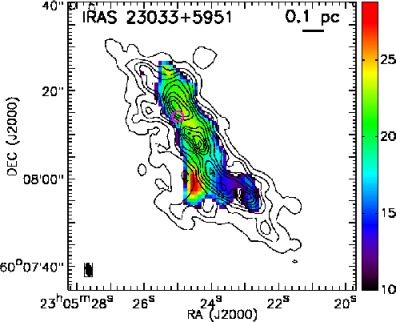} & \includegraphics[width=8.1cm]{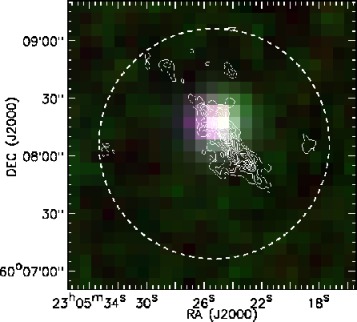} \\
\includegraphics[width=9.0cm]{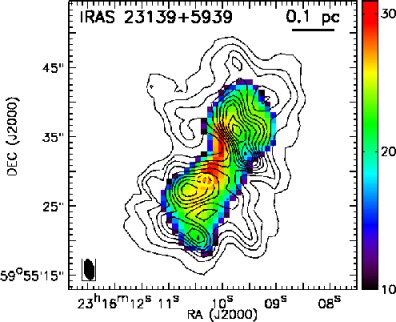} & \includegraphics[width=8.1cm]{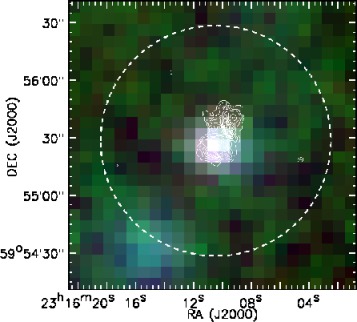} \\
\end{tabular}
\caption{\it Continued}
\end{figure}

\addtocounter{figure}{-1}
\begin{figure}[!t]
\begin{tabular}{p{9.0cm}p{8.1cm}}
\includegraphics[width=8.4cm]{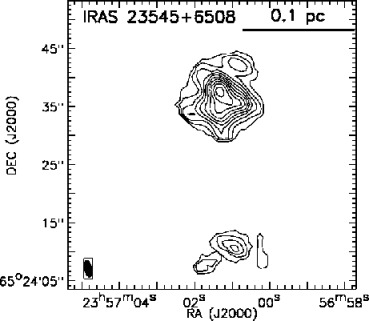} & \includegraphics[width=8.1cm]{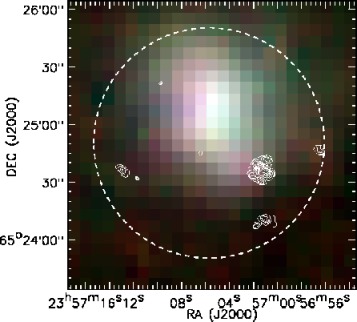} \\
\end{tabular}
\caption{\it Continued}
\end{figure}

\begin{deluxetable}{rccccccccc}
\tabletypesize{\scriptsize}
\tablecaption{Dense core properties.\label{fullcorelist}}
\tablewidth{0pt}
\tablehead{
\multicolumn{1}{c}{\multirow{2}{*}{Core ID}} & RA & Dec & $R$ & $\Delta v$ & $T_R$ & $N(\rm NH_3$) & $M_{core}$ & \multirow{2}{*}{$M_{vir}/M_{core}$} & Multi-Velocity \\
 & (J2000) & (J2000) & (pc) & (km/s) & (K) & (cm$^{-2}$) & (M$_{\odot}$) & & Components
}
\startdata
IRAS 05358+3543 P1 & 05:39:12.27 & +35:45:57.88 & 0.06 & 0.96 & 23.0 &  1.1$\times$10$^{15}$ &    9.6 &  1.1 & \\
P2 & 05:39:12.73 & +35:45:48.29 & 0.04 & 1.66 & 25.9 &  1.2$\times$10$^{15}$ &    4.9 &  4.3 & \\
P3 & 05:39:12.67 & +35:45:37.89 & 0.04 & 1.38 & 21.4 &  9.6$\times$10$^{14}$ &    5.3 &  3.3 & \\
P4 & 05:39:10.56 & +35:46:02.68 & 0.04 & 0.32 & 17.1 &  6.8$\times$10$^{14}$ &    3.5 &  0.2 & \\
P5 & 05:39:13.39 & +35:45:24.29 & 0.02 & 0.64 & 15.0 &  3.4$\times$10$^{14}$ &    0.8 &  2.5 & \\
P6 & 05:39:11.09 & +35:45:05.90 & 0.04 & 0.64 & 15.2 &  3.4$\times$10$^{14}$ &    1.7 &  2.0 & \\
P7 & 05:39:11.81 & +35:45:27.49 & 0.02 & 0.27 & 19.3 &  3.5$\times$10$^{14}$ &    0.6 &  0.6 & \\
P8 & 05:39:13.26 & +35:46:06.68 & 0.02 & 0.76 & 21.3 &  4.3$\times$10$^{14}$ &    1.1 &  2.8 &
\enddata
\tablecomments{Table 6 is published in its entirety in the electronic 
edition of the {\it Astrophysical Journal}.  A portion is shown here 
for guidance regarding its form and content.}
\end{deluxetable}

\end{document}